\documentclass{article}

\usepackage{arxiv}

\usepackage[utf8]{inputenc} 
\usepackage[T1]{fontenc}    
\usepackage{url}            
\usepackage{booktabs}       
\usepackage{amsfonts}       
\usepackage{nicefrac}       
\usepackage{microtype}      
\usepackage{lipsum}		
\usepackage{amssymb,amsmath,amsthm}
\usepackage{listings}
\usepackage{graphicx}
\usepackage{subfig}
\usepackage{apacite}
\usepackage{algorithm}
\usepackage{algorithmicx}
\usepackage{algpseudocode}
\usepackage{kbordermatrix}
\usepackage{todonotes}
\usepackage{natbib}
\usepackage[title]{appendix}

\DeclareMathOperator\supp{supp}

\title{Data assimilation with agent-based models using Markov chain sampling}

\author{
  Daniel Tang\\
    Leeds Institute for Data Analytics, University of Leeds, UK\\
  \texttt{D.Tang@leeds.ac.uk}\\
  \AND
  Nick Malleson\\
  School of Geography, University of Leeds, UK\thanks{This project has received funding from the European Research Council (ERC) under the European Union’s Horizon 2020 research and innovation programme (grant agreement No. 757455)}
  \\  
}

\begin{document}
\maketitle

\begin{abstract}
Every day, weather forecasting centres around the world make use of noisy, incomplete observations of the atmosphere to update their weather forecasts. This process is known as data assimilation, data fusion or state estimation and is best expressed as Bayesian inference: given a set of observations, some prior beliefs and a model of the target system, what is the probability distribution of some set of unobserved quantities or latent variables at some time, possibly in the future?

While data assimilation has developed rapidly in some areas, relatively little progress has been made in performing data assimilation with agent-based models. This has hampered the use of agent-based models to make quantitative claims about real-world systems.

Here we present an algorithm that uses Markov-Chain-Monte-Carlo methods to generate samples of the parameters and trajectories of an agent-based model over a window of time given a set of possibly noisy, aggregated and incomplete observations of the system. This can be used as-is, or as part of a data assimilation cycle or sequential-MCMC algorithm.

Our algorithm is applicable to time-stepping, agent-based models whose agents have a finite set of states and a finite number of ways of acting on the world. As presented the algorithm is only practical for agents with a few bytes of internal state although we discuss ways of removing this restriction. We demonstrate the algorithm by performing data assimilation with an agent-based, spatial predator-prey model.
\end{abstract}


\section{Introduction}

Agent-based models (ABMs) have been widely adopted as an intuitive way to model systems that consist of a heterogeneous collection of autonomous, interacting agents. ABMs are sometimes used to show that an unexpected collective behaviour can emerge from a known agent behaviour.  For example \citet{schelling1971dynamic} showed that agents with only a very slight racial bias can quickly form a highly racially-segregated population. This use of ABMs requires no assimilation of real-world data but generates results that cannot be used to assert anything about the real world without also providing a convincing reason to believe that the model is an abstraction of the real-world. In this paper, we'll deal with an alternative use case in which we start with the following information:
\begin{itemize}
\item A set of real-world observations, $\Omega$, that may contain aggregated or macro-scale quantities and may be subject to noise during the measurement process.

\item An agent-based model expressed as a set of agent behaviours. These will generally be stochastic, in order to account for our uncertainty in agent behaviour, and will be a function of some set of unknown parameters, $\theta$, in order to account for our uncertainty over which behavioural model best represents the real-world target entities. Note that the parameters can, and generally should, contain values that control the size of the model stochasticity in order to account for ``unknown unknowns''. The agents may also interact with a world outside of the model. These ``boundary conditions'' necessarily consist of agents being injected into the model at certain times and/or external influences on modelled agents' behaviour\footnote{we take the purist view that an ABM models everything as an agent}. The start state of the model can also be treated as a boundary condition consisting of the injection of agents into the model at time $t=0$. Although the boundary conditions are semantically distinct from the parameters, their mathematical treatment will be identical so we assume our prior beliefs about boundary conditions is contained in $\theta$. We'll show later how to express an ABM as a ``forecast'' distribution, $P(\tau| \theta)$, which is the probability that a set of agents would exhibit behaviours $\tau$ on a model execution given parameters/boundary conditions $\theta$. We'll call $\tau$ a \textit{model trajectory}.

\item A set of probabilistic prior beliefs, $P(\theta)$, about the model parameters and boundary conditions (after accounting for any micro-calibration or other relevant direct observations we may have). 
\end{itemize}

Our aim will be to calculate the expectation of some (possibly vector-valued) unobserved quantity, $\Lambda$, over the posterior distribution $P(\tau,\theta | \Omega)$
\begin{equation}
\mathbb{E}_{P(\tau,\theta|\Omega)}(\Lambda(\tau,\theta)) = \int \Lambda(\tau,\theta) P(\tau,\theta|\Omega) d\tau d\theta.
\label{expectation}
\end{equation}
Inserting Bayes' rule gives us
\begin{equation}
\mathbb{E}_{P(\tau,\theta|\Omega)}(\Lambda(\tau,\theta)) = \int \Lambda(\tau,\theta) \frac{P(\Omega|\tau)P(\tau|\theta)P(\theta)}{P(\Omega)} d\tau d\theta
\label{bayesassimilation}
\end{equation}
where $P(\Omega|\tau)$ is the observation likelihood, which we assume is easy to calculate given a trajectory that covers the observed period, and $P(\Omega)$ is just a normalising constant which is not usually explicitly calculated when sampling. Note that any information about $P(\tau,\theta | \Omega)$ can be expressed without loss of generality in this way.

There are various approaches in the literature to solving this equation for ABMs. One strategy is to first transform the problem into a simpler form, such as a set of partial differential equations \citep{lloyd_exploring_2016} or a graphical model \citep{liao2010integrated}, for which there exists a well known method of solving the problem. \citet{tang2019data} expresses the problem in terms of creation and annihilation operators then uses symbolic computation to evaluate the integral in \eqref{bayesassimilation}. However, the most popular strategy in the literature is to sample from the posterior, $P(\tau,\theta|\Omega)$, and use the samples to approximate the integral using Monte Carlo integration. This is the technique we employ here.

\subsection{Importance sampling from the prior}
The simplest Monte Carlo approach would be to approximate $\mathbb{E}_{P(\tau,\theta|\Omega)}(\Lambda(\tau,\theta))$ using importance sampling as follows:
\begin{enumerate}
\item Take a sample from the prior parameters/boundary conditions $\theta_i \sim P(\theta)$.
\item Execute the model forward with the given $\theta_i$ to generate a trajectory $\tau_{i} \sim P(\tau|\theta_i)$.
\item Calculate a weight $w_i = P(\Omega|\tau_i)$ to create a weighted sample $\left<\tau_{i},\theta_i, w_i\right>$.
\item Repeat  N times from step 1 to get a set of weighted samples $\left\{\left<\tau_1,\theta_1,w_1\right> \dots \left<\tau_N,\theta_N,w_N\right> \right\}$.
\item Approximate $\mathbb{E}_{P(\tau,\theta|\Omega)}(\Lambda(\tau,\theta)) \approx \frac{1}{\sum_{j=1}^Nw_j}\sum_{i=1}^N w_i\Lambda(\tau_i,\theta_i)$.
\end{enumerate}

However, \citet{chatterjee2018sample} show that the information contained in the weighted samples reduces exponentially with the KL-divergence from the prior to the posterior, $D_{KL}\left(P(\tau,\theta|\Omega) \mid\mid P(\tau|\theta)P(\theta) \right)$. This divergence can be thought of as a measure of the information contained in the observations $\Omega$. For reasons we'll see later, in the case of ABMs, the observations usually contain enough information to make the set of weighted samples less informative than a single unweighted sample, even when $N$ is astronomically high. So this algorithm will not work unless the observations contain very little information.

\subsection{Particle filtering}
In the literature, the most popular way of solving \eqref{bayesassimilation} for ABMs is to use some form of \textit{particle filtering}, also known as \textit{sequential Monte Carlo}. This splits the time period of interest into smaller, contiguous time windows with (not necessarily equidistant) end times $\left<t_1 \dots t_N\right>$ where $t_m < t_{m+1}$. The trajectory and observations are similarly split based on which window they occur in, $\tau = \left<\tau_1 \dots \tau_N\right>$ and $\Omega = \left<\Omega_1 \dots \Omega_N\right>$. The posterior distribution of the whole trajectory can then be split into the product of contributions from each window given the distribution over the previous window, leading to a recursion relation
\begin{equation}
P\left(\tau_{1:t+1}, \theta | \Omega_{1:t+1}\right)
=
\frac{ P(\Omega_{t+1}|\tau_{t+1})
P(\tau_{t+1}|\tau_{1:t},\theta) P\left(\tau_{1:t},\theta| \Omega_{1:t}\right)
}
{	P(\Omega_{t+1}| \Omega_{1:t}) }
\label{bayesrecursion}
\end{equation}
where $\tau_{1:t} = \left<\tau_1 \dots \tau_t\right>$ and $P(\tau_{t+1}|\tau_{1:t},\theta)$ is the probability that an execution of the ABM that starts with trajectory $\tau_{1:t}$ would go on to produce $\tau_{t+1}$ given the parameters/boundary conditions $\theta$. The final posterior $P(\tau_{1:N},\theta|\Omega_{1:N})$ can now be built up in steps, starting with the first window and moving to the last, using the recursion as we go and approximating each $P(\tau_{1:t}, \theta | \Omega_{1:t})$ as a set of samples, or ``particles''. This makes the problem easier in two respects, firstly the dimension of the distributions we need to deal with at each recursion step are reduced since, if we consider each particle separately, we're only concerned with the trajectory in one window at a time, secondly we split the information in the observations into smaller chunks, so at each recursion the information in $\Omega_t$ is more likely to be small enough to make it practical to perform importance sampling from the prior.

If the unobserved quantity of interest, $\Lambda$, depends only on the states of the agents, $\sigma_N$, at time $t_N$ and/or the parameters/boundary conditions, $\theta$, rather than on the full trajectory, $\tau_{1:N}$, then we can reduce the dimensionality even further by performing the recursion on model states rather than trajectories (where a model state tells us how many agents there are in each agent state). First arrange the windows so that each observation lies at the end of a window, then marginalise equation \eqref{bayesrecursion} over $\tau_{1:t+1}$ for a fixed $\sigma_{t+1}$ to get
\begin{equation}
P\left(\sigma_{t+1}, \theta | \Omega_{1:t+1}\right)
=
\frac{ P(\Omega_{t+1}|\sigma_{t+1}) 
}
{	P(\Omega_{t+1}| \Omega_{1:t}) }
\int P(\sigma_{t+1}|\sigma_t,\theta)P\left(\sigma_{t},\theta| \Omega_{1:t}\right) d \sigma_t
\label{bayesstaterecursion}
\end{equation}
where we've assumed that the observation likelihoods depend only on the model state and made explicit in $P(\sigma_{t+1}|\sigma_t,\theta)$ that the probability of the state $\sigma_{t+1}$ depends only on $\sigma_t$ and $\theta$  \footnote{technically, if $\theta$ contains time dependent boundary conditions, the model states need to be timestamped so that the model knows which boundary conditions to apply}.   Note that the definition of ``model state'' can always be re-defined in such a way as to make these assumptions true. In fact, we can see that equation \eqref{bayesrecursion} can be thought of as a special case of equation \eqref{bayesstaterecursion} where the model state, $\sigma_t$, is replaced by the trajectory, $\tau_{1:t}$, in which case the integral over $\sigma_t$ is non-zero for only a single trajectory, since $P(\tau'_{1:t+1}|\tau_{1:t},\theta) = \delta_{\tau_{1:t}}(\tau'_{1:t})P(\tau'_{t+1}|\tau_{1:t},\theta)$, where $\delta$ here is the multivariate delta distribution.

So, we can consider a recursion on \textit{generalised trajectories}, $X_t$,
\begin{equation}
P\left(X_{t+1}| \Omega_{1:t+1}\right)
=
\frac{ P(\Omega_{t+1}|X_{t+1})}
{	P(\Omega_{t+1}| \Omega_{1:t}) }
\int P(X_{t+1}|X_t)P\left(X_{t}| \Omega_{1:t}\right) d X_t
\label{generalisedbayesrecursion}
\end{equation}
so that equations \eqref{bayesrecursion} and \eqref{bayesstaterecursion} would correspond to $X_t=\left<\tau_{1:t},\theta\right>$ and $X_t=\left<\sigma_t,\theta\right>$ respectively.

A particle filter solves this recursion by taking a set of samples $X_t^{1:R}$ drawn from $P(X_t|\Omega_{1:t})$, and generating a set of samples  $X_{t+1}^{1:R}$ drawn from $P(X_{t+1}|\Omega_{1:t+1})$. The simplest algorithm to do this is known as sequential importance resampling which consists of the following steps:
\begin{enumerate}
\item Generate a set of samples $X^{1:R}_0$ from the prior $P(X_0)$. Set $t=0$.

\item For each sample $X_t^i$, generate a sample from the forecast $\hat{X}_{t+1}^i \sim P(X_{t+1}|X^i_t)$ by executing the ABM from the end state of $X^i_t$ to give a set of forecast samples $\hat{X}^{1:R}_{t+1}$.

\item Calculate a weight for each forecast sample $\hat{X}_{t+1}^i$
\[
w_i = \frac{P(\Omega_{t+1}|\hat{X}_{t+1}^i)}{\sum_{j=1}^R P(\Omega_{t+1}|\hat{X}_{t+1}^j)}.
\]

\item Resample from the weighted samples by taking $R$ samples from the approximation
\begin{equation}
P(X_{t+1}|\Omega_{1:t+1}) \approx  \sum_i w_i\delta_{\hat{X}_{t+1}^i}\left(X_{t+1}\right)
\label{importanceApprox}
\end{equation}
to give a new set of unweighted samples $X^{1:R}_{t+1}$. There are a few ways of performing this resampling \citep{douc2005comparison}, the simplest being to draw an integer $i\in[1,R]$ with probability $w_i$, returning the sample $\hat{X}_{t+1}^i$ and repeating $R$ times.

\item If $t<N$, increment $t$ and repeat from step 2.

\end{enumerate}

However, this algorithm suffers from the problem of \textit{sample impoverishment} \citep{li2014fight} or \textit{sample deprivation} which describes the situation when the resampling step leaves many particles in the same state, so the number of distinct samples in our sample set is much smaller than the total number of samples. This is because, as we've seen, after each importance sampling step the information in the observations causes the effective sample size to decrease and so the resampling step is likely to generate repeated samples. This loss of effective sample size can accumulate and in practice as the number of windows increases, the number of particles required to maintain a desired accuracy often increases exponentially. In the worst case, we encounter an observation that is impossible for all particles and the algorithm fails completely. \citet{malleson_simulating_2020} show that sample impoverishment is an issue when applied to data assimilation with an ABM of crowd movement. \citet{khan2003efficient} encounter the same problem in a model of ant movement.

To get a better understanding of impoverishment, suppose we use the above algorithm to generate a set of samples $\left\{\left<\tau_{1:N}^1,\theta^1\right> \dots \left<\tau_{1:N}^S,\theta^S\right>\right\}$ of the full trajectory $P(\tau_{1:N},\theta|\Omega_{1:N})$, from which we generate marginalisations over each window $W_t = \left\{\left<\tau_t^1,\theta^1\right> \dots \left<\tau_t^S,\theta^S\right>\right\}$. On the one hand, the repeated impoverishment at each step causes the number of distinct values in $W_{N-L}$ to decrease exponentially as the lag, L, increases (i.e. as we look further back in time), meaning we lose information about the true distribution of $P(\tau_{N-L},\theta|\Omega_{1:N})$. However, on the other hand, the model dynamics and observations make $P(\tau_{N},\theta|\Omega_{1:N})$ increasingly independent of $P(\tau_{N-L},\theta|\Omega_{1:N})$ as $L$ increases. So if all we care about is the final window, $W_N$, (which is often the case) impoverishment only becomes a problem when the first effect (the rate of loss of information) dominates the second (the rate of increase in independence from the past).
  
Unfortunately every window's trajectory depends on the parameters, $\theta$, but these are sampled in the first window and never subsequently change, so after assimilating only a few windows all samples are likely to have collapsed to the same value of $\theta$, giving us little information about the true distribution of the marginalised posterior $P(\theta|\Omega_{1:N})$. This is fine if the parameters are known at $t=0$, but if there is prior uncertainty in $\theta$ and the observations don't reduce that uncertainty faster than the collapse in information contained in the samples, then impoverishment will be a problem \citep*{liu2001combined, andrieu2004particle}.

A simple way of dealing with sample impoverishment is to add noise to the particles at each step. This is known as roughening \citep*{gordon1993novel, li2014fight} and is equivalent to replacing the delta function in equation \eqref{importanceApprox} with some other function with wider support so we end up with a weighted kernel density estimation. \citet{kieu_dealing_2020} uses this technique in the context of an ABM of public transport. However, \citet{liu2001combined} shows that this leads to over-dispersion unless we ``shrink'' the samples towards their mean value. More generally, the consequence of roughening is that a finite number of samples from a roughened particle filter is no longer a draw from the true posterior, when averaged over all possible draws of the same size, so we need to convince ourselves that the error introduced by the roughening process is small enough for our requirements; not always an easy task for an ABM where distributions can be highly discontinuous.

\citet{wang_data_2015} deals with sample impoverishment in the context of ABM by using the samples at time $t$ to approximate the marginal probabilities of the state of each individual agent at the start of the next window, $P(a_j)$, then drawing new samples from the product of the marginals $P(a_1 \dots a_n) = \prod_{j=1}^NP(a_j)$, i.e. sampling the state of the $j^{th}$ agent from the marginal $P(a_j)$. This improves sample diversity at the price of losing all correlations between agents. 

In the context of ABMs, sample impoverishment can manifest in its most extreme form by generating particles that have zero weight and so contribute no information to the resampling step. This happens because a draw from the forecast of an ABM $P(X_{t+1}|X_t)$ often has an extremely high probability of having a likelihood,$P(\Omega_t|X_{t+1})$ of zero. For example, suppose we're tracking mobile phones by IMEI and receive intermittent signals from the phones identifying which cell they're in. Suppose for simplicity that if a phone is in a known cell at time $t$ then it could be in any of 8 cells at time $t+1$, each with equal prior probability. Suppose at time $t+1$ we get signals from 10 phones. Even if we know the correct model state at time $t$, the probability that a sample from the forecast at time $t+1$ has a non-zero likelihood is $8^{-10}$, about one in a billion. So sequential importance resampling will fail in a single step in this case.

\subsection{Kalman filtering}
One potential solution to the problem of impoverishment is to assume that some or all of the distributions are Gaussian. This is the basis of the data assimilation algorithms that have had a great deal of success in geophysical models \citep*{carrassi2018data, talagrand_assimilation_1997, kalnay_atmospheric_2003, lewis_dynamic_2006}. If the forecast $P(X_{t+1}|\Omega_{1:t})$ and likelihood function $P(\Omega_{t+1}|X_{t+1})$ are multivariate Gaussians, the recursion in \eqref{generalisedbayesrecursion} can be solved analytically and evaluated even when the observations $\Omega_t$ are highly informative. This idea leads to the \textit{Ensemble Kalman filter} \citep{evensen2003ensemble} and the \textit{Unscented Kalman filter} \citep{wan2001unscented}, both of which have been applied to data assimilation in ABM \citep*{ward_dynamic_2016, clay_realtime_2020}.

However, the Gaussian assumption is not usually a good one for ABMs. For example, suppose there is a sports hall containing $N$ people and we take the model state to be the $2N$ dimensional vector consisting of the $(x,y)$ coordinates of each person. Even if the initial state is distributed according to a $2N$ dimensional Gaussian distribution, the either/or decisions of the people can quickly transform this into a multimodal distribution. Suppose two exits are opened on opposite sides of the hall and each person moves towards their nearest exit. Each person makes an either/or decision which exit to move towards based on their position, so we can think of the prior Gaussian as being partitioned into $2^N$ regions corresponding to all possible sets of decisions made by the people. In the forecast distribution, the probability mass in each region will move towards a different model state (one of the $2^N$ states where all agents are at one or other of the exits), creating a highly non-Gaussian distribution with $2^N$ modes. In addition, the likelihood function is often highly non-Gaussian in this state space. For example, suppose there is a sensor at one of the exits that trips if any agents are close. Suppose for example that there are just two agents so the model state space is $(x_1,y_1,x_2,y_2)$ and the origin of the agents' coordinate system is at the sensor. If the sensor trips, the likelihood function is zero everywhere except close to the perpendicular planes $(0,0,x_2,y_2)$ and $(x_1,y_1,0,0)$. This clearly isn't well approximated by any Gaussian. This is a consequence of the fact that we don't know which agent caused the sensor to trip, a problem known as the \textit{data association problem} \citep{lueck_who_2019}. This problem becomes superexponentially worse if we have multiple sensors. If there are $N$ agents and $M \le N$ sensors the likelihood function is the union of $\frac{N!}{(N-M)!}$ hyperplanes.

An alternative representation of the model state space is the \textit{occupation number} representation where each agent state becomes a dimension in the state space and the coordinate values give the number of agents in each state. In this representation Gaussians may be appropriate in some models if there are a large number of agents in every state. However, as the occupation numbers get smaller, an increasing proportion of a Gaussian approximation lies on states that contain negative occupation numbers, which is problematic as a negative occupation number has no natural interpretation. In very high dimensions, the vast majority of the probability mass will lie in a region with at least one negative occupation number. Likelihood functions in this space can also be highly non-Gaussian. For example, the sensor in (a discretised version of) the sports hall example above would have a likelihood function that is a step function on the dimension that corresponds to the state close to the door.

\subsection{Metropolis-Hastings}
In the following sections we present a solution to these problems by using the Metropolis-Hastings algorithm to construct a Markov process whose stationary state is $P(\tau_{1:N},\theta|\Omega_{1:N})$. To our knowledge this is the first time this has been done in a way that is applicable to a wide range of ABMs and observations. For small $N$ we can sample directly from this Markov process to get samples from the posterior. For larger $N$, or if we wish to perform \textit{online} data assimilation where the observations form a continuous, effectively endless stream, we propose using the algorithm as part of a data assimilation cycle or MCMC Particle Filter algorithm \citep*{finke2020limit, septier2009mcmc}. There are many specific ways of implementing this, unfortunately the most appropriate algorithm to use depends on the nature of the ABM dynamics, the likelihood function and the observation operator, $\Lambda$, so there's no silver bullet. We discuss this further in section \ref{discussion} and present some algorithms that are likely to be appropriate in the context of ABMs.

Since the nature of an ABM's parameters is highly model-specific, in the following we focus on the treatment of the trajectory but in a way that is fully integrated with a model-specific treatment of the parameters for which we give more general design advice. In the case of state estimation (where the parameters are already known) the parameter specific part can be left out completely.

\section{Formulation of the problem}

\subsection{Definition of an ABM}
\label{abmdef}
Suppose we have a timestepping ABM that consists of agents with a finite number of possible internal states and a finite number of mutually exclusive ways of acting on their world. Given this, we formally define an ABM as:
\begin{itemize}
	\item An ordered list of agent states $\mathcal{S} = \left<\sigma_0 ... \sigma_{S-1}\right>$.

	\item An ordered list of agent actions $\mathcal{A} =\left< \alpha_0 ... \alpha_{A-1} \right>$.
	
	\item An \textit{agent timestep}, $\pi : \mathbb{Z}\times\mathbb{Z}^S\times\mathbb{Z} \to \mathbb{R}$, which defines the probability that an agent will act in a particular way such that $\pi(\psi,\Psi,a)$ gives the probability that an agent in state $\sigma_\psi$, surrounded by agents, $\Psi$, will perform action $\alpha_a$ (where $\Psi$ is an $S$ dimensional vector whose $i^{th}$ element is the number of agents in state $\sigma_i$ at the start of the timestep).
	
	\item An \textit{action function}, $F: \mathbb{Z} \times \mathbb{Z} \to \mathbb{Z}^S$, which defines the effect of an action on the world such that $F(\psi, a)$ is an $S$ dimensional vector whose $i^{th}$ element gives the number of agents in state $\sigma_i$ that result from an agent in state $\sigma_\psi$ performing act $\alpha_a$ (including the final state of the acting agent).
\end{itemize}

As a simple illustration, we define the ``cat and mouse'' ABM as follows: 
\begin{description}
	\item[Agent states] An agent can be either a cat or a mouse and can be on one of two grid-squares, left or right, so 
	\[
	\mathcal{S} = \left<\textrm{left-cat}, \textrm{right-cat}, \textrm{left-mouse}, \textrm{right-mouse} \right>.
	\]

	\item[Agent actions] In any given timestep, an agent can either move to the other grid-square or stay still, so
	\[
	\mathcal{A} = \left<\textrm{move}, \textrm{stay still}\right>.
	\]
	
	\item[Agent timestep] In a timestep, a cat will move or stay still with probability $0.5$, a mouse will move if there are any cats on the same gridsquare or stay still otherwise. This behaviour can be expressed as the agent timestep:
	\[
	\begin{aligned}
	\pi(\psi, \Psi, a) &=
	\begin{cases}
	0.5 & \text{if } \psi \in \left\{0, 1\right\}\\  
	1 & \text{ if }(\psi = 2, \Psi_1 = 0, a=1) \text{ or } (\psi=3, \Psi_2 = 0, a=1)\\
	1 & \text{ if } (\psi = 2, \Psi_1 > 0, a=0) \text{ or } (\psi=3, \Psi_2 > 0, a=0)\\
	0 & \text{ otherwise.}
	\end{cases}
	\end{aligned}
	\]
	 .

\item[Action function] This gives the result of an agent's actions. For example, $F(1,0)$ is the result of an agent in state $\psi=1$ (\textit{right cat}) performing the action $a=0$ (\textit{move)}, which is one cat in state $\psi=0$ (\textit{left cat}). So $F(1,0) = \{1,0,0,0\}$. This model is simple enough that we can write down $F$ explicitly for all $\psi$ and $a$:
\[
\begin{aligned}
F(0, 0) &= \{0,1,0,0\}\\
F(1, 0) &= \{1,0,0,0\}\\
F(2, 0) &= \{0,0,0,1\}\\
F(3, 0) &= \{0,0,1,0\}\\
F(0, 1) &= \{1,0,0,0\}\\
F(1, 1) &= \{0,1,0,0\}\\
F(2, 1) &= \{0,0,1,0\}\\
F(3, 1) &= \{0,0,0,1\}.\\
\end{aligned}
\]
\end{description}

\subsection{A note on tensor notation}

In the rest of this paper we'll make use of multidimensional arrays. These are just arrays of numbers, like vectors or matrices, but arranged in any number of dimensions. Each array has a \textit{shape} which tells us how many dimensions it has and the number of elements in each dimension. For notational convenience, we'll distinguish between subscript dimensions and superscript dimensions\footnote{covariant and contravariant dimensions if you prefer} and will refer to the set of all arrays of a given shape with the symbol $\mathbb{R}$ adorned with the size of each subscript and superscript dimension.  So, for example, $\mathbb{R}^N_{SA}$ describes the set of all 3-dimensional arrays that have one superscript dimension of size $N$ and two subscript dimensions of sizes $S$ and $A$.

An element of an array is referred to by specifying the sub- and superscript coordinates of the element. For example if $T \in \mathbb{R}^N_{SA}$, then $T^t_{\psi a}$ refers to the element of $T$ that has coordinate $t$ in the superscript dimension and coordinates $(\psi,a)$ in the subscript dimensions. By convention, coordinates begin at 0.

We'll also borrow from tensor notation by using the Einstein summation convention, meaning that if the same index symbol is repeated in sub- and superscript positions in a term, then a summation over that dimension is implied. So, for example if $X \in \mathbb{R}_8$ and $Y \in \mathbb{R}^8$ then
\[
X_\psi Y^\psi \equiv \sum_{\psi=0}^7 X_\psi  Y^\psi.
\]
When the same symbol is repeated in the \textit{same} position with no implied summation, this implies universal quantification. For example if $Y\in R_8$ then
\[
X_\psi = Y_\psi 
\]
is equivalent to
\[
\forall (0 \le \psi < 8) : X_\psi = Y_\psi.
\]
We refer to \textit{slices} of an array by using the $*$ symbol in an index position. For example if $T \in \mathbb{R}^N_{SA}$ then $T^t_{\psi *}$ refers to the 1-dimensional array in $\mathbb{R}_A$ comprised of the elements $\left<T^t_{\psi 0}...T^t_{\psi A-1}\right>$.

The symbols $\mathbf{0}$ and $\mathbf{1}$ represent arrays whose elements are all 0 or 1 respectively. Their shape should be unambiguous from the context.

\subsection{Trajectories}

\begin{figure}
	\centering
	\resizebox{0.5\textwidth}{!}{
		\includegraphics[scale=0.5]{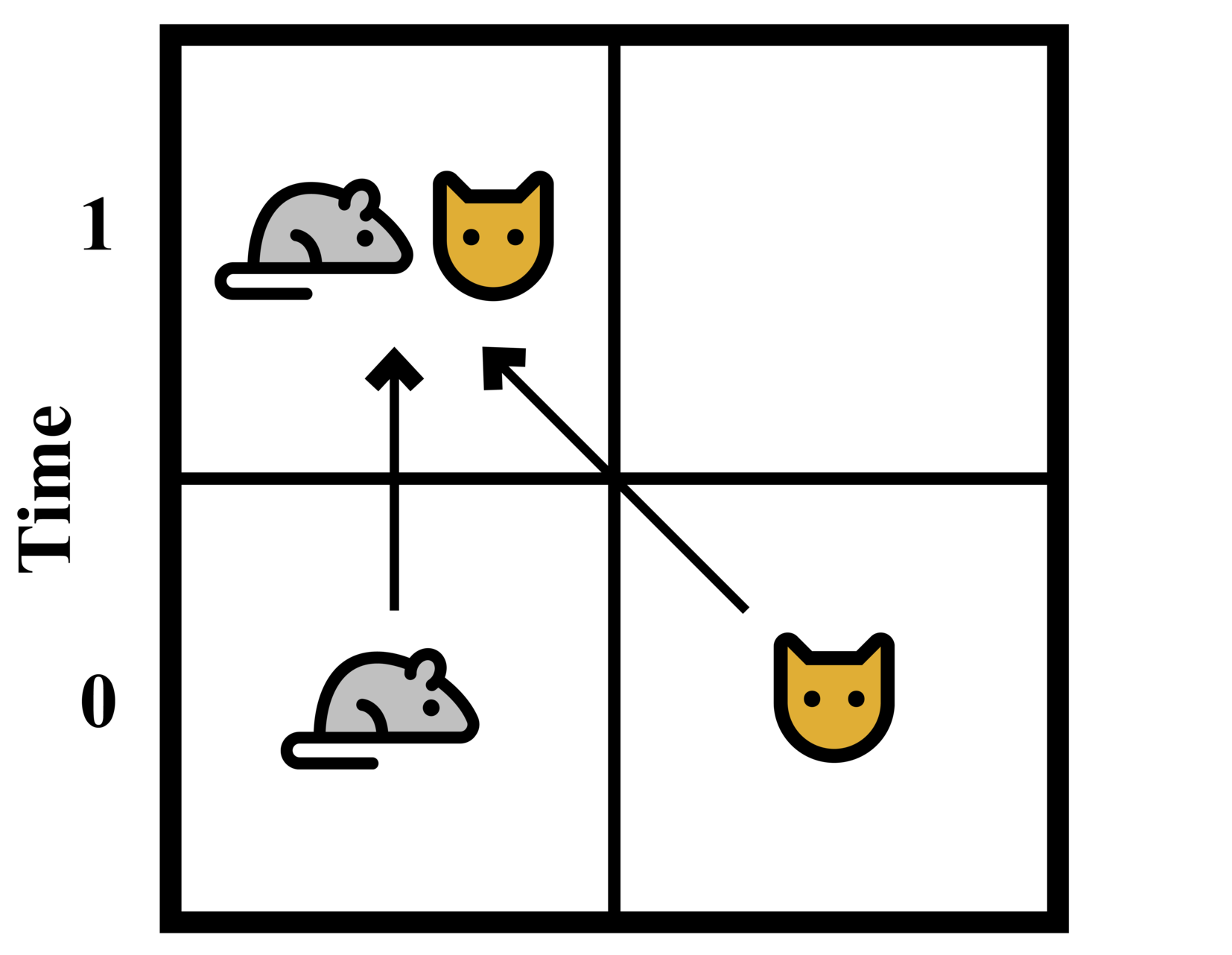}
	}
	\caption{A simple cat and mouse model.\label{fig:AB-MCMC-1}}
\end{figure}

Let a model timestep be an array $E \in \mathbb{R}_{SA}$ whose elements $E_{\psi a}$ give the number of agents in state $\sigma_\psi$ that perform act $\alpha_a$ in this timestep. For example, the timestep shown in Figure~\ref{fig:AB-MCMC-1} for the cat and mouse example would be
\[
E = \kbordermatrix{
	& \alpha_0 & \alpha_1 \\
	\sigma_0 & 0 & 0 \\
	\sigma_1 & 1 & 0 \\
	\sigma_2 & 0  & 1 \\
	\sigma_3 & 0 & 0 \\
}
\]
where one agent in state $\sigma_1$ (\textit{right cat}) performs action $\alpha_0$ (\textit{move}) and one agent in state $\sigma_2$ (\textit{left mouse}) performs action $\alpha_1$ (\textit{stay still}).

Let a model trajectory, $T$, be an array in $\mathbb{R}^N_{SA}$ that represents $N$ timesteps of a model with $S$ agent states and $A$ actions, so that $T^t_{\psi a}$ denotes the $(\psi, a)^{th}$ element of the $t^{th}$ timestep matrix.

An array must satisfy a number of constraints in order to be a valid trajectory of an ABM. Since the elements of a trajectory are counts of agents, they must be non-negative integers. We'll denote the set of all non-negative integer arrays
\begin{equation}
\mathbb{N}^N_{SA} = \left\{ T \in \mathbb{R^N_{SA}}: T^t_{\psi a} \ge \mathbf{0}^t_{\psi a}, T^t_{\psi a} \in \mathbb{Z}\right\}.
\label{nonNegativeInt}
\end{equation}

A trajectory must also be \textit{continuous} by which we mean that any agent that is the result of an action at timestep $t-1$ must be the cause of an action at timestep $t$, and any agents injected by the boundary conditions at timestep $t$ must also act in timestep $t$\footnote{This does not mean that agents cannot leave or enter the system, only that if they do then that change must be the result of an action or boundary condition.}. We call this the \textit{continuity constraint} and define the set of continuous arrays, with respect to an action function $F$:
\begin{equation}
\mathcal{C}^N_{SA}(F) = \left\{T\in\mathbb{R}^N_{SA}: I^0_\phi = T^{0}_{\phi b}\mathbf{1}^b, \forall \left(  1 \le t < N\right): F^{\psi a}_{\phi} T^{t-1}_{\psi a} + I^t_\phi = T^{t}_{\phi b}\mathbf{1}^b  \right\}
\label{continuous}
\end{equation}
where $F \in \mathbb{R}^{SA}_S$ is the action array $F^{\psi a}_* = F(\psi, a)$ and $I^t_\phi$ is the number of boundary condition injections into state $\phi$ at time $t$. Note that this assumes that a timestep is performed by updating all agents in parallel, i.e. agents must act with no information about the actions of other agents in the same timestep. This is different from serial update, where agents are updated in a particular order within a timestep and have access to information about the actions of agents that come before them in the ordering. Note also that this assumes that agents in the same state are indistinguishable, i.e. if we swap two agents in the same state, the trajectory is unchanged.

So, we define the set of trajectories, $\mathcal{T}^N_{SA}(F)$, as set of arrays that satisfy \eqref{nonNegativeInt} and \eqref{continuous}.
\begin{equation}
\mathcal{T}^N_{SA}(F) = \mathbb{N}^N_{SA} \cap \mathcal{C}^N_{SA}(F).
\label{SetOfTrajectories}
\end{equation}

\subsection{The posterior of an ABM}

Our goal is to generate samples from the Bayesian posterior over the trajectories, boundary conditions and model parameters, given a set of observations, $\Omega$
\[
P\left(T,\theta \middle| \Omega\right) \propto P\left(\Omega \middle| T\right)P(T|\theta)P(\theta).
\]
The model forecast, $P(T|\theta)$, can be decomposed into the product of conditionals for each timestep and agent state
\[
P\left(T \middle| \theta \right) \propto  P(\Psi^0_*|\theta) \prod_{t,\psi} P(T^t_{\psi *} | \Psi^t_*,\psi,\theta)
\]
where $\Psi^t_* \in\mathbb{N}_S$ is the model state at the start of timestep $t$ and $P(\Psi^0_*|\theta)$ is defined by the boundary conditions\footnote{for notational clarity we assume that all injections are at time $t=0$, but this can easily be extended to injections at any or all timesteps.}.  The term $P(T^t_{\psi *} | \Psi^t_*,\psi,\theta)$ is the probability that $\Psi^t_\psi$ agents starting in state $\psi$ will collectively perform the vector of actions $T^t_{\psi *}$. The probability that a single agent performs action $a$ is given by the agent timestep $\pi_\theta(\psi,\Psi^t,a)$ (where we've added the subscript $\theta$ to emphasise its possible dependence on the parameters). Since agent actions are simultaneous, the collective behaviour is given by the multinomial distribution
\begin{equation}
P\left(T^t_{\psi *} \mid \Psi^t_*, \psi, \theta\right) = 
\begin{cases}
\Psi^t_\psi!\prod_a \frac{\pi_\theta(\psi,\Psi^t_*,a)^{T^t_{\psi a}}}{T^t_{\psi a}!} & \text{ if } T^t_{\psi a}\mathbf{1}^a = \Psi^t_\psi \\
0 & \text{otherwise.}
\end{cases}
\end{equation}
If the trajectory is continuous then $\Psi^t_* = T^t_{* a}\mathbf{1}^a$ so the model forecast is given by 
\begin{equation}
P(T|\theta) =
\begin{cases}
P(\Psi^0_* = T^0_{* c}\mathbf{1}^c|\theta)
\prod_{t, \psi} \left(T^t_{\psi b} \mathbf{1}^b \right)!
\prod_{a} \frac{\pi_\theta(\psi, T^{t}_{* d}\mathbf{1}^d,a)^{T^{t}_{\psi a}}}{T^{t}_{\psi a}!} & \text{if } T \in \mathcal{T}^N_{SA}(F) \\
0 & \text{otherwise.}\\
\end{cases}
\label{priorTrajectory}
\end{equation}

The likelihood function, $P(\Omega|T)$, can also usually be decomposed. Without loss of generality, we take $\Omega$ to consist of some number of observations that are independent of each other given the trajectory, so that $\Omega$ is a set of pairs $(\omega,v)$ that consist of a stochastic observation operator $\omega$ and an observed value $v$ (which may be a vector). We write $P(\omega(T)=v)$ to denote the probability of observation operator $\omega$ making observation $v$ on trajectory $T$. So
\[
P(\Omega|T) = \prod_{(\omega,v) \in \Omega} P(\omega(T)=v)
\]
and the posterior can be written as
\begin{equation}
P(T,\theta|\Omega) \propto 
\begin{cases}
P(\theta)
P(\Psi^0_* = T^0_{* c}\mathbf{1}^c|\theta)
\prod_{(\omega,v) \in \Omega}
P\left(\omega(T)=v\right)
\prod_{t, \psi} \left(T^t_{\psi b} \mathbf{1}^b \right)!
\prod_{a} \frac{\pi_\theta(\psi, T^{t}_{* d}\mathbf{1}^d,a)^{T^{t}_{\psi a}}}{T^{t}_{\psi a}!} & 
 \text{if } T \in \mathcal{T}^N_{SA}(F) \\
0 & \text{otherwise.}\\
\end{cases}
\label{posterior}
\end{equation}

\section{Approximating the support of the posterior}

As we saw in the introduction, sampling from equation \eqref{posterior} is difficult because it is often very hard to find proposal trajectories that have non-zero posterior probability. Our strategy to solve this problem is to first derive an expression for a volume of trajectory/parameter space that bounds the support of the posterior, $\supp(P(T,\theta|\Omega))$ (i.e. the set of $\left<T,\theta\right>$ pairs that have non-zero posterior probability). If we choose a bounding volume that it is easy to sample from and is a relatively tight approximation of $\supp(P(T|\Omega)))$ then there's a good chance that a sample drawn from this volume will have non-zero probability. The bounding volume doesn't need to be exact, it just needs to be tight enough to give us a good chance of finding a non-zero probability sample.

From equation \eqref{posterior}
\begin{equation}
\begin{aligned}
\supp (P( T,\theta |\Omega)) = 
& \bigcap_{(\omega,v) \in \Omega,t, \psi, a} \mathcal{T}^N_{SA} \cap \supp(P(\theta)) \\ 
&\supp(P(\Psi^0_* = T^0_{* c}\mathbf{1}^c)|\theta) \cap \\
& \supp\left(P\left(\omega(T)=v\right)\right) \cap \\
&\left( \supp\left(\pi_\theta(\psi,T^t_{* b}\mathbf{1}^b,a)\right) \cup \left\{T:T^t_{\psi a} = 0\right\} \right)
\end{aligned}
\label{support}
\end{equation}
i.e. in order for $\left<T,\theta\right>$ to have non-zero posterior probability, $\theta$ must have non-zero prior probability and $T$ must be a trajectory of the ABM that has a possible start state, all observations must have non-zero likelihood and each non-zero element of $T$ must denote an agent action with non-zero probability.

\subsection{Convex mixed-integer polyhedra}
\label{BPoly}

We now approximate equation \eqref{support} with a \textit{convex, mixed-integer polyhedron}, $\mathcal{P}$, (see, for example, \citet{conforti2010polyhedral}) which we define to be a set of vectors, some of whose elements are integer and some real-valued, that satisfy a set of linear inequalities:
\[
\mathcal{P} = \left\{ X\in\mathbb{Z}^N \times \mathbb{R}^M : L \le  CX \le U \right\}
\]
for some matrix $C$ and some vectors $L$ and $U$ .

Any linear inequality on $T$ and $\theta$ can be expressed in this form by letting $X$ consist of the elements of $T$ and $\theta$ flattened into one dimension so that
\[
X^i = W^{\psi a i}_{t}T^t_{\psi a} + G^i_j\theta^j
\]
and $W$ and $G$ have inverses
\[
\hat{W}^t_{\psi a i} = W^{\psi' a' i'}_{t'}\delta_{\psi'\psi}\delta_{a'a}\delta^{t't}\delta_{i'i}
\]
and
\[
\hat{G}^j_i = G^{i'}_{j'}\delta_{i'i}\delta^{j'j}
\]
such that
\[
T^t_{\psi a} = \hat{W}^{t}_{\psi a i}X^i
\]
and
\[
\theta^j = \hat{G}^{j}_{i} X^i
\]
where $\delta^{ij}$ and $\delta_{ij}$ are 1 if $i=j$ and 0 otherwise. In this way we can switch unambiguously between $X$ and $\left<T,\theta\right>$.

We now consider how to express each term in equation \eqref{support} as a mixed integer polyhedron. From equation \eqref{SetOfTrajectories} we can see immediately that the set of all trajectories, $\mathcal{T}^N_{SA}$, is is already defined as a set of linear constraints on $T$. The support of the prior $P(\theta)$ can usually be easily bounded by a hyper-rectangle giving the range of each variable. The supports of the start state, $P(\Psi^0_*|\theta)$, the observations, $P(\omega(T)=v)$, and the agent actions, $\pi_\theta(\psi,T^t_{*b}\mathbf{1}^b,a)$, can often be easily expressed as linear constraints. However, if this is not the case, each of the probability distributions can be expressed as a computer program. In practice, these computer programs will be simple and it will be possible to use a technique known as \textit{abstract interpretation} \citep{cousot1977abstract} using the domain of convex polyhedra \citep*{cousot1978automatic, becchi2018efficient, fukuda2020polyhedral} to efficiently calculate a convex polyhedron that bounds the support of the program. Software to perform abstract interpretation using convex polyhedra already exists \citep*{henry2012pagai, GN2021, jeannet2009apron, bagnara2008parma} and the technique has been used in applications such as compiler optimization \citep{nsjodin2009design} and verification of safety-critical systems \citep{halbwachs1997verification}. The programs may contain calls to a random number generator \texttt{Random()} that returns a random floating-point number $0 \le r < 1$. This can be represented in the polyhedral domain by introducing an real valued auxiliary variable, $r$, that satisfies $0 \le r < 1$ for each call to \texttt{Random()}. These auxiliary variables can be removed immediately using the abstract interpretation software to project the polyhedron back into the space of $X$. This relies on converting the polyhedron to its set of vertices \citep{motzkin1953double} and projecting the vertices, so the number of vertices must be of a tractable size. Alternatively, the auxiliary variables can be added to X and marginalised out when we take the samples.

If the number of agents is very much smaller than the number of agent states (which is often the case with agent-based models) then we may be willing to make the assumption that in any timestep there is at most one agent performing a given action from a given start state (i.e. $\forall \psi, a, t: T^{\psi a}_t \in \{0,1\}$). Under this assumption, which we'll call the \textit{Fermionic assumption}, the \textit{Fermionic trajectories}, $\mathcal{F}^N_{SA} = \left\{T\in\mathcal{T}^N_{SA}: \forall \psi, a, t: T^{\psi a}_t \in \left\{0,1\right\}\right\}$, all lie on the corners of the unit hypercube. So the intersection of $\mathcal{F}^N_{SA}$ with any set is a convex polyhedron and $\supp(P(T|\Omega))$ can always be exactly represented as a mixed-integer polyhedron (finding that polyhedron isn't always tractable, though, in which case abstract interpretation software can be used to find a larger polyhedron that contains it).

So, if we let $\mathcal{P}(f)$ be a mixed integer polyhedron that contains $\supp(f)$ then from equation \eqref{support} we can approximate the support of the posterior as
\begin{equation}
\begin{aligned}
\supp(P( T, \theta |\Omega)) \subseteq \mathcal{P}(P(W^{\psi a i}_{t}T^t_{\psi a} + G^i_j\theta^j|\Omega)) =
& \bigcap_{(\omega,v) \in \Omega,t,\psi, a} \mathcal{T}^N_{SA} \cap \mathcal{P}(P(\theta)) \\
& \mathcal{P}(P(\Psi^0_* = T^0_{* c}\mathbf{1}^c|\theta)) \cap\\
&    \mathcal{P}\left(P\left(\omega(T)=v\right)\right) \cap \\
& 
\left(\mathcal{P}\left(\pi_\theta(\psi,T^t_{* b}\mathbf{1}^b,a)\right)
\cup
\left\{T: T^t_{\psi a} = 0\right\}\right).
\\
\end{aligned}
\label{polyhedralSupport}
\end{equation}

The intersection of two mixed-integer polyhedra is easy to express as another mixed integer polyhedron by just concatenating the constraints
\begin{multline}
\left\{ X\in\mathbb{Z}^N \times \mathbb{R}^M : L \le  CX \le U \right\}
\cap
\left\{ X\in\mathbb{Z}^{N} \times \mathbb{R}^{M} : L' \le  C'X \le U' \right\}\\
= \left\{ X\in\mathbb{Z}^{N} \times \mathbb{R}^{M} : {L \choose L'}  \le   {C \choose C'}X \le {U \choose U'} \right\}
\label{intersection}
\end{multline}
so the only difficulty in calculating $\mathcal{P}(P(X|\Omega))$ is the union in the final term of equation \eqref{polyhedralSupport}. We transform this into an intersection by introducing an auxiliary variable $z$ and using the following identity: if the hyper-rectangle $0 \le X \le H$ bounds $L \le C X \le U$ for some vector $H \in \mathbb{Z}^N \times \mathbb{R}^M$, then for any given $i$ 
\begin{multline}
\left\{ X\in\mathbb{Z}^N \times \mathbb{R}^M : L \le C X \le U \right\}
\cup
\left\{X: X^i = 0, \mathbf{0} \le X \le H\right\}
= \left\{ \right.X\in\mathbb{Z}^N \times \mathbb{R}^M, z\in\{0,1\}:\\
CX + (\overline{B}-U)z \le \overline{B},\\
\underline{B} \le CX + (\underline{B}-L)z,\\
0 \le H^iz - X^i,\\
 z - X^i \le 0 \left. \right\}\\
\\
\label{implication}
\end{multline}
where $\overline{B}$ are upper bounds on the values of $CX$ in the hyper-rectangle, defined as
\[
\overline{B} = \frac{\left(C+\left|C\right|\right)H}{2}
\]
where $|C|$ here denotes an element-wise absolute value of a matrix and $\underline{B}$ are lower bounds on the values of $CX$ defined as
\[
\underline{B} = \frac{\left(C - \left|C\right|\right)H}{2}.
\]

To see why this identity holds, note first that the constraints make $z$ into an indicator variable that is 0 if $X^i=0$ or 1 otherwise. When $z=1$ the first constraint on the right hand side of \eqref{implication} is equal to $CX \le U$ and the second is equal to $L \le CX$ so their intersection is $L \le  CX \le U$ as required, whereas when $z=0$ the constraints are equal to $\underline{B} \le CX \le \overline{B}$. But $\underline{B}$ and $\overline{B}$ are lower and upper bounds on the values of $CX$ in the hyper-rectangle so this is satisfied for all values $0 \le X \le H$ as required. 

There are two things worth noting here. Firstly if we make the Fermionic assumption then $z = X^i = T^t_{\psi a}$ so the auxiliary indicator variable becomes unnecessary. Secondly, we must bound $\mathcal{P}(\pi_\theta(\psi,T^t_{*b}\mathbf{1}^b,a))$ by a hyper-rectangle with its lower corner at the origin. In practice, this is not a problem as we can just make a change of variable, imposing upper and lower bounds on each variable such that the probability of being outside those bounds is negligible, then shifting if the lower bound isn't on zero.

So, the support of the posterior can be approximated by a mixed-integer polyhedron, $\mathcal{P}(P(X|\Omega))$, by using equations \eqref{polyhedralSupport}, \eqref{intersection} and \eqref{implication}.

As a simple illustration, consider a two-timestep trajectory of the cat and mouse model described in section \ref{abmdef}. Suppose we flip a fair coin to decide whether each agent state is occupied or empty at $t=0$. Suppose also that we observe a cat in the left grid-square at time $t=1$. Our aim is to construct a polyhedron, $\mathcal{P}(P(X|\Omega))$, that bounds the support of the posterior.

Working through \eqref{polyhedralSupport} term by term, the $\mathcal{T}^2_{4\,2}$ term constrains to non-negative, continuous trajectories as defined in equation \eqref{SetOfTrajectories}, which is already in linear form. The second term is the support of the prior over the parameters which we ignore here since there are no parameters. The next term is the support of the start state, since we're flipping a coin for each agent state at $t=0$, each state occupation must be at most 1, which can be expressed as
\[
\left\{T:T^0_{\psi 0} + T^0_{\psi 1} \le \mathbf{1}_{\psi}\right\}.
\]
for each $\psi$. The fourth term is the support of the observation. Since we observe a cat in the left grid-square at time $t=1$ we add the constraint
\[
T^1_{0 0} + T^1_{0 1} = 1.
\]
The final term is the constraint due to agent interactions. The impossible interactions are a mouse staying put when there is a cat on the same gridsquare or moving when there are no cats, so we constrain against these
\begin{equation}
\begin{aligned}
\supp(\pi(2,T^t_{* a}\mathbf{1}^a,0)) &= \left\{ T: -T^t_{0 0} - T^t_{0 1} \le -1 \right\}\\
\supp(\pi(3,T^t_{* a}\mathbf{1}^a,0)) &= \left\{ T: -T^t_{1 0} - T^t_{1 1} \le -1 \right\}\\
\supp(\pi(2,T^t_{* a}\mathbf{1}^a,1)) &= \left\{ T: T^t_{0 0} + T^t_{0 1} \le 0 \right\}\\
\supp(\pi(3,T^t_{* a}\mathbf{1}^a,1)) &= \left\{ T: T^t_{1 0} + T^t_{1 1} \le 0 \right\}
\end{aligned}
\label{actionConstraints}
\end{equation}
for all t. If, for simplicity, we make the Fermionic assumption by adding the constraints
\[
\mathbf{0}^t_{\psi a} \le T^t_{\psi a} \le \mathbf{1}^t_{\psi a}
\]
then using the identity in \eqref{implication} to take the union of each constraint in \eqref{actionConstraints} with $\left\{T: T^t_{\psi a} = 0, \mathbf{0} \le T \le \mathbf{1}\right\}$ gives the four constraints
\[
\begin{aligned}
-T^t_{0 0} - T^t_{0 1} + T^t_{2 0} & \le 0\\
-T^t_{1 0} - T^t_{1 1} + T^t_{3 0} & \le 0\\
T^t_{0 0} + T^t_{0 1} + 2T^t_{2 1} & \le 2 \\
T^t_{1 0} + T^t_{1 1} + 2T^t_{3 1} & \le 2
\end{aligned}
\]
for each timestep $t=0$ and $t=1$.

Taken together, these constraints define a polyhedron that bounds the set of (Fermionic) trajectories for the cat and mouse ABM with the given start state distribution and observation.

\section{Sampling from the posterior}
\label{samplingFromThePosterior}
Having shown how to calculate $\mathcal{P}(P(X|\Omega))$, we now show how to use this to sample from $P(X|\Omega)$.

Various methods of sampling from a mixed-integer polyhedron exist in the literature. \citet{baumert2009discrete} presents a very general algorithm for sampling from any weighted subset of integer points within a bounding hyper-rectangle. The algorithm relies on sampling points on a random walk. However, in our case the number of valid trajectories could be a vanishingly small proportion of the number of points in the smallest bounding hyper-rectangle, so a random walk would be unlikely to come across any valid trajectories. Universal hashing \citep{meel2016constrained} provides a promising way of sampling uniformly from the integer points in a convex polyhedron, but we have found that this technique doesn't scale to the number of dimensions needed for our application.

The Metropolis-Hastings algorithm is another well known and widely used sampling algorithm and this is the approach we take here. To do this we need to define
\begin{itemize}
\item a set of Markov states, $\mathcal{M}$, along with a mapping, $E:\mathcal{M} \to \mathbb{Z}^N \times \mathbb{R}^M$, which maps Markov states to sample values, $X$,

\item a stochastic proposal function $f:\mathcal{M} \to \mathcal{M}$ from which we can generate transitions between Markov states,

\item a probability distribution $P: \mathcal{M} \to \mathbb{R}$ which gives the probability of each Markov state (this need not be normalised as the Metropolis Hastings algorithm only ever needs probability ratios).

\end{itemize}

In order to be of use in practice, the proposal function, $f$, must have the following properties:
\begin{itemize}
	\item For any two Markov states there should exist a sequence of transitions which forms a path between those states and has a non-zero probability of being proposed and accepted.
	
	\item Given a current Markov state, there should be a computationally efficient procedure to generate a proposal and calculate its acceptance probability. 
\end{itemize}

\subsection{The set of Markov states}

Given the polyhedron $\mathcal{P}(P(X|\Omega))$, we split the constraints into two sets: equalities (i.e. those whose lower and upper bound have the same value) and inequalities (i.e. those whose lower and upper bounds differ) so that
\begin{equation}
\mathcal{P}(P(X|\Omega)) = \left\{X \in \mathbb{Z}^N \times \mathbb{R}^M: \left(L \le CX \le U\right) \cap \left(DX = E\right) \right\}.
\label{zPolySupport}
\end{equation}
Our aim will be to use the equality constraints to reduce the dimension of the problem and so improve the efficiency of the algorithm. Note that $D$ has at least $(N-1)S$ rows since the continuity constraints \eqref{continuous} are equality constraints. 

Suppose there are $N_e$ equality constraints and we partition the elements of $X$ into `basic' and `non-basic' elements, $X_B$ and $X_N$ respectively, so that there are exactly $N_e$ basic elements.

If we let $Q$ be a permutation matrix that separates the basic from non-basic elements so that
\[
QX = {X_B \choose X_N}
\]
then we can write
\[
DX = DQ^{-1}{X_B \choose X_N} = \left(B \mid N\right){X_B \choose X_N} = E
\]
so
\begin{equation}
BX_B + NX_N = E.
\label{eqconstraints}
\end{equation}

Now, since there are $N_e$ basic variables and $N_e$ equality constraints, $B$ is square so if we choose the basic variables to ensure that $B$ has an inverse then
\begin{equation}
X_B = B^{-1}(E - NX_N)
\label{basicvars}
\end{equation}
so letting
\[
V = Q^{-1}{B^{-1}E \choose \mathbf{0}}, \, M = Q^{-1}{-B^{-1}N \choose \mathbf{1}}
\]
gives
\begin{equation}
X = V + MX_N.
\label{markovtotrajectory}
\end{equation}
In section \ref{basis} we'll show how to choose $X_B$ to ensure that $B$ has an inverse and the integer elements of $X_B$ remain integer as long as the integer elements of $X_N$ are integer. Given this, then the original polyhedron is equivalent to the reduced dimension polyhedron
\begin{equation}
\mathcal{P}' = \left\{X_N \in \mathbb{Z}^{N'} \times \mathbb{R}^{M'}: L-CV \le  CMX_N \le U-CV\right\}
\label{reducedPolySupport}
\end{equation}
where $N'$ and $M'$ are the number of integer and real-valued elements remaining in $X_N$.

Since $Q$ is just a permutation matrix, $X_N$ and $X_B$ can be bound within hyper-rectangles by transforming the $\mathcal{P}$-bounding hyper-rectangle, $H$, from section \ref{BPoly}
\[
QH = {H_B \choose H_N}.
\]
So, $X_N$ is bound by $0 \le X_N \le H_N$. Given this, we define the set of Markov states to be
\[
\mathcal{M} = \left\{ X_N: 0 \le X_N \le H_N, \right\}.
\]
and equation \eqref{markovtotrajectory} maps each Markov state, $X_N \in \mathcal{M}$, to a unique sample, $X$. Although this set includes values outside of the polyhedron \eqref{reducedPolySupport}, we'll show a way to deal with that later.

\subsection{The proposal function}
\label{theProposalFunction}
In order to propose a new Markov state given a current state, $X_N$, first choose an element of $X_N$ and a direction in which to perturb that element. This defines a \textit{rounded perturbed state}, $X'_N$,  which is equal to $X_N$ but with one of its elements rounded to the nearest integer and perturbed by $\pm 1$. Only rounded perturbed states that remain inside the hyper-rectangle $H_N$ are considered. Real valued variables should be scaled so that a perturbation of $\pm 1$ usually changes $P(X|\Omega)$ by a small amount. The boundaries, $H_N$, should be chosen to lie on an integer value plus 0.5 for real-valued variables. So, we can define the set of all valid perturbations
\[
\mathcal{D}(X_N) = \left\{\left<i,\delta\right> : X'_N = \lfloor X_N \rceil_i + \delta e_i, 0 \le X'_N \le H_N, \delta \in \left\{+1,-1\right\}, i \in \mathcal{V} \right\}
\]
where $e_i$ is the unit vector with a 1 in the $i^{th}$ element, $\mathcal{V}$ is the set of indices of $X_N$ and we use the notation $\lfloor . \rceil_i$ to denote the rounding of the $i^{th}$ element of a vector to the nearest integer.

Assume for now that we have an approximation of the target distribution $\tilde{P}(X_N) \approx P(X_N|\Omega)$. Given a current state, $X_N$, the probability of choosing perturbation $\left<i,\delta\right> \in \mathcal{D}(X_N)$ is given by
\begin{equation}
P(X_N,i,\delta) = \frac{\min\left(1, \frac{\tilde{P}(\lfloor X_N \rceil_i + \delta e_i)}{\tilde{P}(\lfloor X_N \rceil_i)}\right)}{S(X_N)} 
\label{transitionProb}
\end{equation}
where
\begin{equation}
S(X_N) = \sum_{\left<i,\delta\right> \in \mathcal{D}(X_N)} \min\left(1, \frac{\tilde{P}(\lfloor X_N \rceil_i + \delta e_i)}{\tilde{P}(\lfloor X_N \rceil_i)}\right)
\label{transitionSum}
\end{equation}

Given a perturbation $\left<i,\delta\right>$, the final proposed state is given by
\[
X'_N = \lfloor X_N \rceil_i + (\delta + \epsilon)e_i
\]
where $\epsilon$ is a \textit{fractional perturbation} drawn with uniform probability from $[-0.5:0.5]$ if the $i^{th}$ element is real-valued or $\epsilon = 0$ if it is integer-valued.

\subsubsection{Approximating the posterior}
\label{approximatingThePosterior}
The proposal function requires an approximation of the target distribution, $P(X|\Omega)$. For this, we use a function of the form
\begin{equation}
\tilde{P}(X) = \prod_i \tilde{P}_i(Z_iX)
\end{equation}
where $Z_i$ are row vectors and $\tilde{P}_i$ are univariate factors of any form. We call this a \textit{linearly factorized distribution}, where the linearity lies in the $Z_iX$ terms while the $\tilde{P}_i$ can be non-linear.

The first thing to note is that we can express $\mathcal{P}(P(X|\Omega))$ as a linearly factorized distribution in the following way: first define the infeasibility function $\iota(l,x,u)$ to be equal to 0 if $l < x < u$ or equal to the distance to the nearest bound otherwise
\[
\iota(l,x,u) =
\begin{cases}
x-u & \text{if }x>u\\
l-x & \text{if }x<l\\
0 & \text{otherwise.}
\end{cases}
\]
Now, given a set of constraints $L \le CX \le U$, if we define a factor for each constraint
\begin{equation}
\tilde{P}_i(C_iX) = e^{\frac{-\iota(L_i,C_iX,U_i)}{\tau}}
\end{equation}
then
\begin{equation}
\prod_i \tilde{P}_i(C_iX) = e^{\frac{-\sum_i \iota(L_i,C_iX,U_i)}{\tau}}
\label{lfconstraints}
\end{equation}
is a linearly factorized distribution that is 1 inside the polyhedron and that decays exponentially outside at a rate set by $\tau$. So as $\tau \to 0$ equation \eqref{lfconstraints} tends to an indicator function for the set of points in $\mathcal{P}(P(X|\Omega))$.

A linearly factorized distribution in $X$ can easily be transformed into a linearly factorized distribution in $X_N$ by using the linear transform in equation \eqref{markovtotrajectory}. Transformation to $X_N$ ensures that the resulting distribution satisfies the equality constraints $D(V+MX_N)=E$, while the factors in equation \eqref{lfconstraints} can be used to ensure that $0 \le X_B \le H_B$ and that the constraints in \eqref{reducedPolySupport} are satisfied.

Turning our attention now to the distribution inside the polyhedron, if we look back at equation \eqref{posterior} we can see that the posterior of an ABM is already highly factorized and many of the factors are already linearly factorised.

However, when considering the efficiency of the final algorithm, there is a trade off between the closeness of $\tilde{P}(X_N)$ to $P(X_N|\Omega)$ and the number of factors involved in its calculation. In our experiments, we have found we can attain very good acceptance probabilities with relatively simple renderings of $\tilde{P}(X_N)$. The start state, likelihood function and prior over $\theta$ are often already uncorrelated functions of the elements of $\theta$ and the state occupation numbers, so are already in the correct form. This leaves only the multinomial term in equation \eqref{posterior} which we approximate with the linearly factorized
\begin{equation}
\prod_{t, \psi}\left(T^t_{\psi b}\mathbf{1}^b\right)! \prod_a \frac{\tilde{\pi}^t_{\psi a}\,^{T^t_{\psi a}}}{T^t_{\psi a}!}
\label{multinomialapprox}
\end{equation}
where
\[
\log(\tilde{\pi}^t_{\psi a}) = \frac{\int T^t_{\psi a}\log\left(\pi_\theta(\psi, \Psi^t, a)\right)P(T,\theta) dT d\theta}{\int T^t_{\psi a}P(T,\theta) dT d\theta}.
\]
This integral can be calculated by drawing samples from the prior $P(T,\theta)$ and using Monte Carlo integration. This approximation marginalises over $\theta$, but if $\pi_\theta$ has a particularly strong dependence on $\theta$ then it may be worth capturing some of that dependence by, for example, replacing the $T^t_{\psi a}$ term in \eqref{multinomialapprox} with $T^t_{\psi a} + R\theta$ for some row vector $R$. The $\tilde{\pi}^t_{\psi a}$ can then be calculated by minimising the expectation of the difference between the log probabilities of the normalised approximation and the true multinomial.

So, the final approximation, $\tilde{P}(X_N)$, is given by the product of the factors of the bounding convex polyhedron \eqref{lfconstraints}, the multinomial approximation \eqref{multinomialapprox} and the approximations of the observations, the start state (boundary conditions) and $P(\theta)$.

Since some of the Markov states are infeasible, calculating $\tilde{P}$ for these states may require the evaluation of a factor at a value that is outside its domain (for example, equation \eqref{multinomialapprox} is undefined if $T^t_{\psi a}$ is negative). In this case we choose the value at the nearest point that is within the domain.

\subsection{The probability of a Markov state}
\label{probabilityOfAMarkovState}
If a Markov state satisfies all the constraints in \eqref{reducedPolySupport} then its probability is defined to be equal to the true posterior given by \eqref{posterior}
\[
P(X_N) = P(X|\Omega) = P(T,\theta|\Omega)
\]
where
\[
X=V + MX_N = W^{\psi a i}_{t}T^t_{\psi a} + G^i_j\theta^j.
\]
If the constraints are not all satisfied, then the probability is defined to be the product of the factors in \eqref{posterior} and the infeasibility factors in \eqref{lfconstraints}. If any factor in \eqref{posterior} becomes zero as a consequence of the infeasibility, then it is replaced by its linearly factorized approximation (which must be non-zero otherwise the state wouldn't have been proposed).

This means that, as long as $\tau$ is finite, the Markov chain will pass through some infeasible states. However, if we just ignore these and only take samples from feasible states then we end up with samples from the true posterior. Allowing the Markov chain to pass through infeasible states has the advantage of ensuring that there exists a path between any two feasible states and improves mixing. When used in the context of a particle filter, it also solves the problem of unexpected observations that can't be generated by any particle, as we can allow the particles to become temporarily infeasible. The price we pay for this is the computational cost of moving through infeasible states that don't generate useful samples.

It's worth noting that equation \eqref{lfconstraints} is a Boltzmann distribution which describes a thermodynamic system whose ``energy'' is equal to the infeasibility of $X$ that is at equilibrium with a heat bath of ``temperature'' $\tau$. It has been shown that simulations of thermodynamic systems of this type are able to solve large sets of linear integer inequalities like the one we're dealing with here \citep{kirkpatrick1983optimization}. Here, we won't be changing the temperature during the sampling process, but we do need to choose a value for $\tau$. Higher temperatures will increase mixing in the Markov chain, but will also increase the proportion of time spent in infeasible states so we need to find a temperature that is high enough to ensure good mixing of the chain and low enough to ensure that a reasonable proportion of samples are feasible. We have found that these two competing effects roughly cancel each other out when measuring the overall efficiency of the algorithm in effective samples per second, so the value of $\tau$ is not critical as long as it's within a certain range. We have found that a good rule of thumb is to set the temperature so that 50-80\% of the samples in a chain are infeasible.

From the Metropolis-Hastings algorithm, the probability of accepting a proposal $X'_N = \lfloor X_N \rceil_i + (\delta + \epsilon)e_i$ is given by
\[
\alpha = 
\min\left( 1, \frac{P(X'_N)P(X'_N, i, -\delta)}{P(X_N)P(X_N,i,\delta)} \right) =
\min\left(1, \frac{P(X'_N)}{P(X_N)} \frac{\tilde{P}(\lfloor X_N \rceil_i)}{\tilde{P}(\lfloor X'_N \rceil_i)}  \frac{S(X_N)}{S(X'_N)}\right).
\]
Since $\tilde{P}(X_N)$ approximates $P(X_N)$, $\frac{P(X'_N)}{P(X_N)} \frac{\tilde{P}(\lfloor X_N \rceil_i)}{\tilde{P}(\lfloor X'_N \rceil_i)}$ should be close to 1. We'll see in the next section that if we choose the basis carefully then $S(X_N)$ will not change very much between transitions so the ratio $ \frac{S(X_N)}{S(X'_N)}$ should also be close to 1 meaning that a good proportion of proposals should be accepted. In our experiments we achieved acceptance rates of around $85\%$.

\subsection{Choosing a basis}
\label{basis}
Our definition of a Markov state depends on a partition of $X$ into basic and non-basic variables such that $B$ in equation \eqref{basicvars} has an inverse, $B^{-1}$, and if $X_N \in \mathbb{Z}^{N'} \times \mathbb{R}^{M'}$ then $X \in \mathbb{Z}^{N} \times \mathbb{R}^{M}$ (i.e. if $X_N$ has all integers in the correct elements then so does $X$). In general there exist many partitions that satisfy these requirements so it remains to define a method of choosing one. The choice of basis affects the efficiency of the Markov chain via:
\begin{itemize}
\item the computational cost of updating the proposal probabilities $P(X_N,i,\delta)$ and the sum $S(X_N)$ after each transition.

\item the effect of the ratio of sums $S(X_N)/S(X'_N)$ on the acceptance probability

\item the mixing rate of the Markov chain, since the choice of basis affects the available transitions in $X$ space.
\end{itemize}

From \eqref{transitionProb} and \eqref{transitionSum} we can see that in order to calculate $P(X_N,i,\delta)$ and $S(X_N)$ we need to know $\frac{\tilde{P}(\lfloor X_N \rceil_i + \delta e_i)}{\tilde{P}(\lfloor X_N \rceil_i)}$ for all $\left<i,\delta\right> \in \mathcal{D}(X_N)$.  Since $\tilde{P}$ is linearly factorised
\[
\frac{\tilde{P}(\lfloor X_N \rceil_i + \delta e_i)}{\tilde{P}(\lfloor X_N \rceil_i)} =
\frac{\prod_j \tilde{P}_j(Z_j(\lfloor X_N \rceil_i + \delta e_i))}{\prod_j \tilde{P}_j(Z_j\lfloor X_N \rceil_i)} =
\prod_{j : Z_j e_i \ne 0} \frac{\tilde{P}_j(Z_j \lfloor X_N \rceil_i + \delta Z_j e_i)}{\tilde{P}_j(Z_j \lfloor X_N \rceil_i)}.
\]
Let
\[
P'(X_N,i,\delta) = \min\left(1, 
\prod_{j : Z_je_i \ne 0} \frac{\tilde{P}_j(Z_j \lfloor X_N \rceil_i + \delta Z_j e_i))}{\tilde{P}_j(Z_j \lfloor X_N \rceil_i)}\right)
\]
be the un-normalised proposal probability. If we accept a state transition $\left<k,\delta'\right>$, $P'(X_N,i,\delta)$ needs to be updated 
to $P'(\lfloor X_N \rceil_k + (\delta'+\epsilon)e_k, i, \delta)$. However, this update only affects factors $\tilde{P}_j$ such that  $\left\{j : Z_j e_k \ne 0 \right\}$, so a transition $\left<k,\delta'\right>$ only changes $P'(X_N,i,\delta)$ if $\left\{j : Z_je_i \ne 0\right\}$ intersects with $\left\{j : Z_je_k \ne 0\right\}$. So, if we let $Z$ be the matrix formed from the row vectors $Z_i$ and choose a basis such that $Z$ is sparse, $P'(X_N,i,\delta)$ can be stored for each value of $\left<i,\delta\right> \in \mathcal{D}(X_N)$, and only values for a few values of $i$ will need to be recalculated on transition. \citet{TangMutableCategorical} shows that a binary sum-tree data structure can be used to efficiently update and draw from the un-normalised probability distribution defined by $P'(X_N,i,\delta)$ while also updating $S(X_N)$.
 
The effect of $\frac{S(X_N)}{S(X'_N)}$ on the acceptance probability is also improved if $Z$ is sparse because only a small proportion of the terms in the sum change between $S(X_N)$ and $S(X'_N)$ so we should expect their ratio to be close to 1.

The Markov chain will also mix better if the columns of $Z$ are sparse. \citet{mihelich2018maximum} shows that the Kolmogorov-Sinai entropy of the Markov process is a good proxy for its mixing speed. In the context of our algorithm the entropy is higher if the neighbours of a Markov state have similar probability. If $Z$ is sparse then adjacent Markov states are more likely to have similar probabilities, for a fixed mean element magnitude.

So, our aim will be to find a basis that keeps $Z$ sparse, while maintaining the correct integer signature of $X$. To do this we use the following algorithm: Start with a linearly factorized distribution $\tilde{P}$ on $X$ (rather than $X_N$)
\[
\tilde{P}(X) = \prod_i \tilde{P}_i(C_iX).
\]
and a set of equality constraints that need to be satisfied for all Markov states $X_N$
\[
DX = E.
\]
From these, form a mixed-integer polyhedron
\[
\mathcal{P} = \left\{X \in \mathbb{Z}^N \times \mathbb{R}^M: L \le CX \le U \cap DX = E \right\}.
\]
We assume, without loss of generality, that all constraints involving only integer elements of $X$ have integer coefficients. $C$ is usually sparse to start with, so our aim is to maintain that sparsity while using the equality constraints to reduce $X$ to $X_N$. If we let
\[
W= {D \choose C}
\]
and
\[
F = {E \choose \mathbf{0}}
\]
then
\begin{equation}
\mathcal{P} = \left\{X \in \mathbb{Z}^N \times \mathbb{R}^M: {\mathbf{0} \choose L} \le W X - F \le {\mathbf{0} \choose U} \right\}
\label{jointconstraints}
\end{equation}

Our aim is to mark variables from $X$ as basic one at a time. Once we're done, the remaining variables will make up the non-basic variables, $X_N$.

Given a set of constraints in the form \eqref{jointconstraints}, we mark the $j^{th}$ element of $X$ as basic and the $i^{th}$ equality constraint as ``reduced'' by choosing an element $W^i_j$ to be a \textit{pivot point} and performing Gaussian elimination on the $j^{th}$ column of $W$ so that $W^i_j$ is the only non-zero element in the column $W_j$. This can be done by pre-multiplying by $G$
\[
{\mathbf{0} \choose L} \le GWX - GF \le {\mathbf{0} \choose U}
\]
where $G$ is the matrix
\[
G =  
\begin{pmatrix}
1 &  & &-\frac{W^0_j}{W^i_j} & & &\\
& \ddots & &\vdots & & &\\
& & 1 & -\frac{W^{i-1}_j}{W^i_j} & & &\\
& & & \frac{1}{W^i_j} & & &\\
& &  & -\frac{W^{i+1}_j}{W^i_j} & 1 & &\\
& & & \vdots & & \ddots &\\
& & & -\frac{W^n_j}{W^i_j} & & &1\\
\end{pmatrix}
\]

Letting $\hat{W} = GW$ and $\hat{F} = GF$ recovers the canonical form
\begin{equation}
{\mathbf{0} \choose L} \le \hat{W}X - \hat{F} \le {\mathbf{0} \choose U}
\label{eliminatedConstraint}
\end{equation}
but now the $j^{th}$ element of $X$ is marked as basic and the $i^{th}$ row of $\hat{W}$ is marked as ``reduced''. We now repeat the process, choosing only pivot points that are non-basic and unreduced, until no more equality constraints can be reduced.

The use of Gaussian elimination here ensures that $B^{-1}$ exists. In order to ensure that $X_B$ has the correct integer structure, given that $X_N$ does, we add the requirement that we can only mark integer elements of $X$ as basic on a pivot point $W^i_j$ such that the row vector $W^i$ is zero on all non-integer columns and $W^i_j$ divides $W^i_k$ for all $k$ (so for rows that have a non-zero element on a real-valued column, we always eliminate on a real-valued column). This requirement ensures that for integer elements, $X_B^i$, after reduction $X_B^i = F^i - W^iX_N$. But if $X_N$ has the correct integer structure then $W^iX_N$ is a sum of products of integers, which is an integer. So, if $X_B^i$ has any integer solution then $F^i$ must also be an integer. So, if $X_N$ has the correct integer structure, then $X_B^i$ is also an integer.

This defines the set of valid pivot points at each elimination step. In order to decide which point to choose, we note that when we pivot on $W^i_j$, $GW$ differs from $W$ in at most 
\[
\mu = (\left|W_j\right|_0-1)(\left|W^i\right|_0-1)
\]
 elements, where $\left|\cdot\right|_0$ is the L0-norm (i.e. the number of non-zero elements in a vector or matrix). So, this gives an upper bound on $\left|WG\right|_0 - \left|G\right|_0$, the change in L0-norm (or reduction in sparsity) of $W$ when we pivot on $W^i_j$. It has been found that choosing $W^i_j$ to be the element that minimises $\mu$ (known as the Markowitz criterion) is a computationally efficient way of maintaining the sparsity of a matrix while performing Gaussian elimination \citep*{markowitz1957elimination, suhl1990computing, maros2002computational}. So, in order to find a sparse basis, we calculate the set of valid pivot points, choose the point that minimises $\mu$, perform the pivot and repeat until no more valid pivot points remain. \citet{suhl1990computing} presents a computationally efficient algorithm for doing this.

\section{Spatial predator-prey demonstration}

We demonstrate the above techniques on a 32x32 grid of ``predator'' and ``prey'' agents. The agent acts consist of moving to an adjacent grid-square (i.e. up, down, left or right), giving birth to another agent on the an adjacent grid-square or dying. If any agent moves off the edge of the grid, it reappears at the opposite edge (i.e. the grid is topologically a torus). Predators only give birth when there is at least one prey on an adjacent grid-square (i.e. if there is a source of food close by). Prey may be eaten by predators so their probability of dying is much greater when predators are present on adjacent grid-squares. The probability of each agent's behaviour is shown in table \ref{rates}. These values were chosen to minimise the probability that either species becomes extinct.
\begin{table}
	\begin{center}
		\begin{tabular}{llllc}
		\hline
		Aget type & Adjacent agents & Behaviour & Probability\\
		\hline
		Prey & No predators & die &        0.100\\
			& & give birth &        0.156\\
			& & move &        0.744\\
			& &&\\
		Prey & Predators > 0 & die &        0.400\\
			& & give birth &        0.156\\
			& & move &        0.444\\
			& &&\\
		Predator  & No prey & die  &      0.100\\
			& & give birth &        0.000\\
			& & move &        0.900\\
			& &&\\
		Predator  & Prey > 0 & die  &      0.100\\
			& & give birth &        0.300\\
			& & move &        0.600\\
		\hline& 
		\end{tabular}
	\end{center}
	\caption{The probabilities of each behaviour in the predator-prey model}
	\label{rates}
\end{table}

In order to generate observations to assimilate, we used the following procedure:
\begin{enumerate}
\item Generate a start state by drawing the number of predators and prey in each gridsquare at $t=0$ from a Bernoulli distribution with probability $0.05$. i.e. for $\Psi^0_* \in \{0,1\}^S$
\begin{equation}
P(\Psi^0_*) = \prod_\psi 0.05^{\Psi^0_\psi}0.95^{1-\Psi^0_\psi}
\label{bernoulliStartState}
\end{equation}

\item Simulate the ABM for 16 timesteps from the start state. If the resulting trajectory isn't Fermionic, repeat from step 1. The resulting trajectory, $T_{\text{real}}$, is considered to be the ``real-world'' trajectory from which we take observations.

\item For each timestep and gridsquare of $T_{\text{real}}$, take two draws from a Bernoulli distribution with probability $0.05$. If the first draw is 1, then observe the number of predators in that gridsquare at that time. If the second draw is 1 then observe the number of Prey. The observations were noiseless so the observed count was the true count.
\end{enumerate}

Given the (Fermionic) model, the start state $P(\Psi^0_*)$ and the observations, an approximation of the posterior was generated as described in section \ref{approximatingThePosterior} and used to generate a sparse basis as described in section \ref{basis}. The resulting basis had 22.4 non-zero elements per million.

Four initial feasible solutions were generated in the following way:
\begin{enumerate}
\item Draw a start state from $P(\Psi^0_*)$.
\item Simulate for 16 timesteps to get a (non-Fermionic) prior trajectory.
\item Starting with the prior trajectory, generate proposals as described in section \ref{theProposalFunction} and accept with probability 1.0 until a feasible state is reached.
\end{enumerate}

Starting with these four initial solutions, four separate Markov chains were generated using the Matropolis-Hastings algorithm as described in section \ref{samplingFromThePosterior}. 1,000,000 samples were then taken from each chain and discarded in order to burn them in. Finally, 10,000,000 samples were taken from each chain and split into first and last halves to give 8 sample sequences of 5,000,000 samples each. For all samples, the temperature, $\tau$, was set to $0.1$.

Feasible samples were generated at a rate of one every $42\mu$s on a single core of a 1.6GHz Intel i5-8250U. $85\%$ of the proposals were accepted and $70\%$ of samples were infeasible. 

\begin{figure}
	\centering
	\resizebox{0.5\textwidth}{!}{
		\includegraphics[scale=0.5]{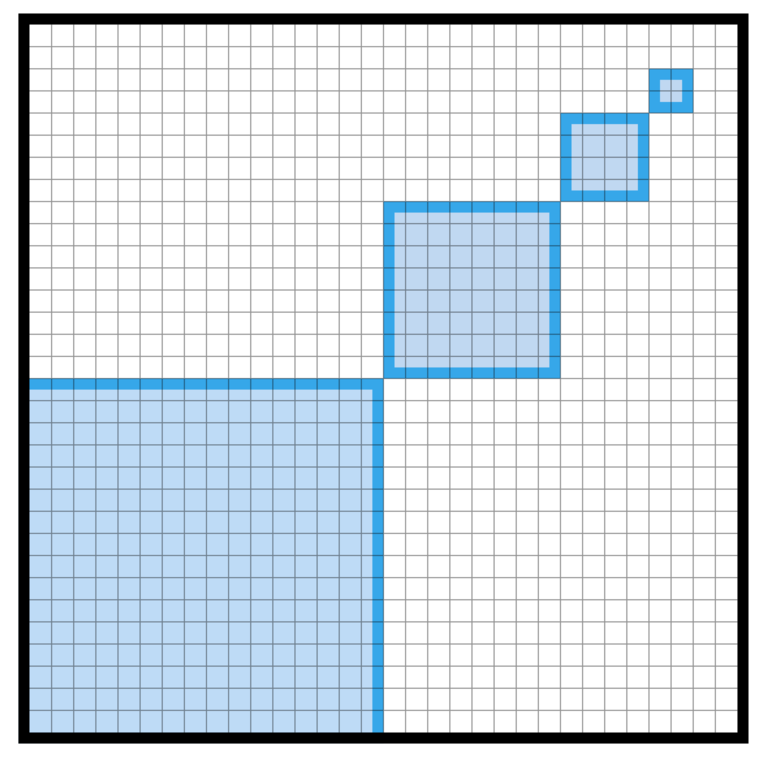}
	}
	\caption{Summary statistics consist of the total number of agents in each of the four shaded regions.}
	\label{figRegions}
\end{figure}
In order to assess the convergence of the chains, a set of summary statistics were calculated for each sample of each sequence. The summary statistics consisted of the number of agents within each shaded square shown in figure \ref{figRegions} measured at the end of the last timestep. We'll refer to these as statistic 1 to 4 going from largest to smallest area. This arrangement of regions was chosen in order to capture the convergence at different spatial scales. From the summary statistics, we calculated the the Gelman-Rubin diagnostic \citep{gelman1992inference}. This gives a measure of the uncertainty in the standard deviation of the statistic due to the finite length of the sample sequences. If this number is close to 1, then we have some justification in believing that we have taken enough samples. A value of less than 1.1 is often used as a criterion for convergence. We also approximated the autocorrelation of each statistic at various time lags for each sample sequence. The autocorrelation of a convergent chain should decline to zero long before the lag reaches the total number of samples. Finally, we approximated the number of effective samples in each sequence (i.e. the number of samples from a perfect IID sampler that would give the same uncertainty in the mean of the statistic). Details on how these values were calculated are given in the appendix and our results are shown in figure \ref{figAutocorrelation} and table \ref{convergenceResults}.

\begin{figure}
	\centering
	\resizebox{0.9\textwidth}{!}{
		\includegraphics{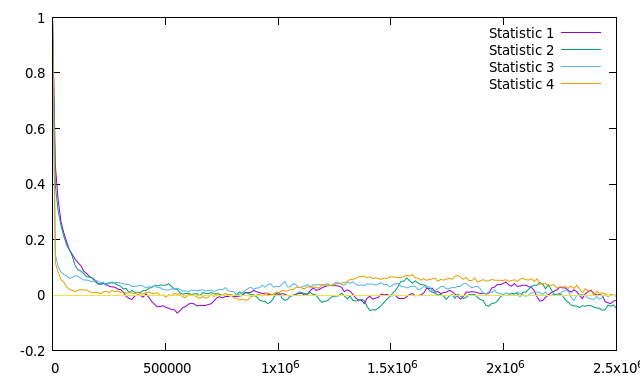}
	}
	\caption{Mean autocorrelation for each of the summary statistics, averaged over all chains, as a function of the lag in number of samples.}
	\label{figAutocorrelation}
\end{figure}

\begin{table}
	\begin{center}
		\begin{tabular}{llllc}
			\hline
			Measure & Statistic 1 & Statistic 2 & Statistic 3 & Statistic 4\\
			\hline
			Gelman-Rubin & 1.00727 & 1.01167 & 1.00451 & 1.00153  \\
			Effective samples & 420 & 417 & 489 & 1019 \\
			\hline
		\end{tabular}
	\end{center}
	\caption{The Gelman Rubin diagnostic and number of effective samples per sample-sequence for each statistic}
	\label{convergenceResults}
\end{table}

\begin{figure}
	\centering
	\resizebox{0.9\textwidth}{!}{
		\includegraphics{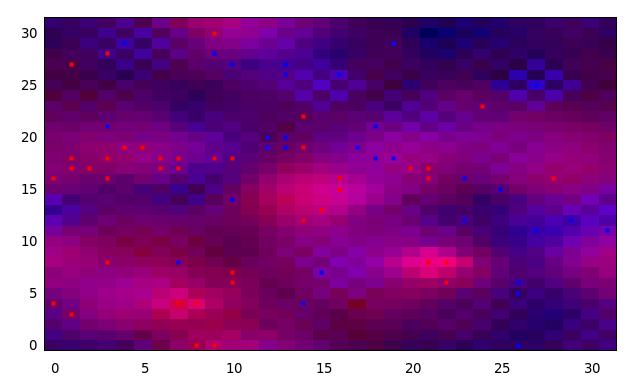}
	}
	\caption{A representation of the positions of the agents at time $t=16$. The dots represent the positions of agents in the ``real'' trajectory, $T_{\text{real}}$, from which the observations were generated. Blue dots represent predators, red dots represent prey. The background colour of each square has blue/red intensity proportional to the log of the mean number of predators/prey in that square averaged over all samples.}
	\label{figEndState}
\end{figure}

Finally, we take our unobserved values, $\Lambda$, to be the expected number of agents in each state at the end of the simulation. Figure \ref{figEndState} shows this averaged over all samples, along with the end state of the ``real'' trajectory $T_{\text{real}}$. This gives a visual demonstration that the observations have given rise to a posterior that provides information about the true positions of the agents.

The code used to generate these results can be found at \url{https://github.com/danftang/AgentBasedMCMC}.

\section{Discussion} 
\label{discussion}

\subsection{Sequential MCMC for ABM}

The predator-prey example demonstrated assimilation over a relatively short time period. In practice it's likely that assimilation would be required over a longer period, or perhaps over a continuous stream of observations. As discussed earlier, the way to deal with this is to split the longer time period into windows as in equation \eqref{generalisedbayesrecursion} and treat each window separately. We can make use of the MCMC sampler to implement this broad idea in many different ways.

One way would be the following: given $N$ samples from the posterior of a window $X^{1:N}_{t} \sim P(X_{t}|\Omega_{1:t})$, approximate the posterior from some family of simpler, analytical distributions, $P_\Theta(X_t)$, by finding the $\Theta$ that minimises the KL-divergence from $P_\Theta(X_t)$ to $P(X_{t}|\Omega_{1:t})$. The KL-divergence can itself be approximated from the samples $X^{1:N}_t$ with
\[
D_{KL}\left(P(X_{t}|\Omega_{1:t}) \mid P_\Theta(X_t)  \right) \approx
\sum_{s=1}^N \frac{1}{N}\log\left(\frac{1}{N P_\Theta(X^s_t)}\right)
\]
which is minimised when
\[
\Theta \approx \min_\Theta \left(-\sum_{s=1}^N \log\left(P_\Theta\left(X^s_t\right)\right) \right).
\]
The approximation, $P_\Theta(X_t)$, can then be inserted into the recursion equation \eqref{generalisedbayesrecursion} to generate samples from the next window $X^{1:N}_{t+1} \sim P(X_{t+1}|\Omega_{1:t+1})$, from where we can repeat the cycle indefinitely. The exact form of $P_\Theta$ is model dependent. The simplest form would be a product of univariate Poisson distributions, one for each of the integer variables, and Gaussians for the real-valued variables. If we wish to capture correlations between variables, a multivariate Poisson distribution could be used (see for example \citet{munoz2021multivariate}, where $P_\Theta$ is a product of univariate Poisson distributions whose rate parameters are given by $\Lambda = B X_R + A X_I$, where $X_R$ are the real-valued variables and $X_I$ are the integer valued variables, $B$ and $A$ are matrices, and $A$ is triangular).

An alternative approach is to use the \textit{Resample-Move} algorithm presented in \citet{gilks2001following}. This is a particle filter based on the sequential importance resampling algorithm described earlier. The innovation comes in the form of an additional MCMC \textit{move} step after each resample step. At each move step each particle, $X^i_t = \left<\tau_{1:t}^i,\theta^i\right>$, is ``moved'' by initialising the Markov chain with $X^i_t$ and drawing a single new sample $X'^{i}_{t} \sim P(X_t|\Omega_{1:t})$ by taking one or more Markov transitions. This solves the problem of particle impoverishment because even if multiple samples collapse to the same state during the resample step, the subsequent move step is likely to move them apart again. The algorithm remains exact because the samples coming out of the $m^{th}$ resample step are already distributed according to the target distribution, $P(X_t|\Omega_{1:t})$, but this is also the stationary distribution of the Markov transition so the moved samples are also distributed according to the target distribution.

The Resample-Move algorithm still uses an importance sampling step with a draw from the prior which, as we have seen, is often a problem for ABMs because drawing from the prior is likely to result in a sample with zero weight. If this is the case, we can replace the draw from the prior with an draw from $P(X_{t+1}|\Omega_{1:{t+1}},X^i_t)$ using our MCMC algorithm for each particle $X^i_t$ from the previous timestep. The correct weight of the particle is then proportional to $P(\Omega_{t+1}|X^i_t)$ \citep{doucet2009tutorial}. Unfortunately we don't immediately know this value. If we're doing an assimilation step every timestep, or if the ABM is simple enough, it may be possible to calculate this value analytically, however, often this is not possible.   \citet*{han2001markov, newton1994approximate, stefankovic2009adaptive} give a number of ways of approximating these values, at least up to a multiplicative constant (which is all we need). An alternative is to maintain a population of feasible and infeasible particles. At each window, draw $X_t$ from the prior, extract $X_N$ to give a Markov state then weight by the Markov probability as defined in \ref{probabilityOfAMarkovState} divided by the prior probability of drawing $X_N$. The set of feasible particles are then draws from the posterior.

As the number of assimilated windows increases, there may be a lag $L$ where we are happy to assume that the new observations $\Omega_{t+1}$ do not change our beliefs about the model state before the end of the $(t-L)^{th}$ window, i.e. $P(\tau_{1:t-L}|\Omega_{1:t},\theta) \approx P(\tau_{1:t-L}|\Omega_{1:t+1},\theta)$. In this case, it is sufficient to perturb the trajectories of only a finite number of windows into the past when performing the MCMC moves, so we perturb $\tau_{t-L+1:t+1}$ and $\theta$ while holding $\tau_{1:t-L}$ fixed. This means that the dimension of the MCMC problem does not increase as we assimilate more windows. The early parts of the trajectory are then subject to possible particle impoverishment again if we're using an algorithm with a resampling step, but this time it doesn't matter because it is, by assumption, not affecting the distribution of the current state.

\subsection{Limitations and further work}
This algorithm presented here is intended as the proof-of-concept of a novel approach to MCMC sampling from an ABM rather than a production-ready tool for immediate use by practitioners. At present, the bounding of the supports of the agent time-step, boundary conditions and observation likelihood functions by a convex polyhedron must either be done by hand or by third party abstract interpretation tools. Code for commonly encountered scenarios is available on the Github repository but this process could be automated for user supplied functions by taking as input a computer program and outputting a set of linear constraints for the support. Similarly, the temperature setting, $\tau$, at present needs to be set by trial-and-error, whereas an automated warm-up or adaptive scheme would improve the algorithm's efficiency and usability.

As presented, the algorithm is only applicable if the internal state of an agent can be represented in a few bytes. This is because it depends on explicitly representing the probability of each Markov transition at each iteration. But the number of possible transitions is of the order of the number of non-basic variables, and this expands exponentially with the size of the internal state of an agent. So, if agents have a large amount of internal state it will become impractical to represent this distribution explicitly. This problem can be overcome by considering only a subset of non-basic variables for update at each iteration, in a similar way to \textit{partial pricing} in the simplex algorithm \citep{maros2002computational}. For example, given the current Markov state, only consider updating the non-basic variables that correspond to alternative actions of some agent in the current trajectory at some timestep. In this situation we also can't explicitly store the coefficients of the linear constraints, $C$, so in order to calculate the effect of a perturbation on the basic variables we'd calculate this on the fly from equations \eqref{nonNegativeInt} and \eqref{continuous}, the agent timestep function and the observation operators. Finding a good basis of non-basic variables could also be done on the fly for some subset of variables, or alternatively equality constraints could be left as-is as inequalities with equal lower and upper bounds. More work needs to be done exploring the relative efficiency of these options.

\section{Conclusion}

We have described and demonstrated an algorithm to sample model-trajectories and parameters from the posterior distribution of an agent-based model given a set of observations. We derived an analytical expression for the posterior and showed how to generate a convex polyhedron that bounds the support of this distribution. We then showed how these can be used to construct an approximation of the posterior in the form of a linearly factorized distribution, and how this can be embedded in a Metropolis-Hastings algorithm to generate samples from the posterior. This allowed us to perform Bayesian inference with a spatial predator/prey model and so demonstrate data assimilation on an ABM which would be impossible with other techniques described in the literature.

The current lack of tools to perform data assimilation with agent-based models has restricted their use in data analytics. The development of more powerful techniques for performing Bayesian inference with ABM could transform the usefulness and applicability of these models and the algorithm described here will hopefully be a first step towards this transformation.

\begin{appendices} 

\section{Calculation of convergence statistics}
\label{calculatingConvergenceStats}
Given $m$ sample sequences, each containing $n$ samples, let $x_{ij}$ to refer to the $i^{th}$ sample of the $j^{th}$ sequence and let
\[
\bar{x}_j = \frac{1}{n}\sum_{i=1}^n x_{ij}
\]
be the mean of the $j^{th}$ sequence and
\[
\bar{x} = \frac{1}{m}\sum_{j=1}^m \bar{x}_j
\]
be the mean over all sequences.

Let $W$ be the within-sequence variance
\[
W = \frac{1}{m} \sum_{j=1}^m \frac{1}{n-1} \sum_{i=1}^n (x_{ij} - \bar{x}_j)^2
\]
and $B$ be the between-sequence variance
\[
B = \frac{n}{m-1}\sum_{j=1}^m (\bar{x}_j - \bar{x})^2
\]
Following \citet{gelman2013bayesian}, an overapproximation of the true variance of the statistic can be calculated as
\[
\widehat{\text{var}}^+ = \frac{n-1}{n}W + \frac{1}{n}B
\]
and the Gelman-Rubin statistic can be defined as
\[
\hat{R} = \sqrt{\frac{\widehat{\text{var}}^+}{W}}
\]

Also following \citet{gelman2013bayesian}, we approximate the autocorrelation as
\[
\hat{\rho}_t = 1 - \frac{V_t}{2\widehat{\text{var}}^+}
\]
where
\[
V_t = \frac{1}{n-t} \sum_{i=1}^{n-t} (x_i - x_{i+t})^2.
\]

$\hat{\rho}_t$ can be used to calculate the effective number of samples using
\[
\hat{n}_e = \frac{mn}{1 + 2\sum_{t} \hat{\rho}_t}
\]
where the sum runs from $t=0$ until the first $t$ such that $\hat{\rho}_t \le 0$.

\end{appendices}

\bibliographystyle{apacite}
\bibliography{references}

\begin{thebibliography}{}

\bibitem [\protect \citeauthoryear {%
Andrieu%
, Doucet%
, Singh%
\BCBL {}\ \BBA {} Tadic%
}{%
Andrieu%
\ \protect \BOthers {.}}{%
{\protect \APACyear {2004}}%
}]{%
andrieu2004particle}
\APACinsertmetastar {%
andrieu2004particle}%
\begin{APACrefauthors}%
Andrieu, C.%
, Doucet, A.%
, Singh, S\BPBI S.%
\BCBL {}\ \BBA {} Tadic, V\BPBI B.%
\end{APACrefauthors}%
\unskip\
\newblock
\APACrefYearMonthDay{2004}{}{}.
\newblock
{\BBOQ}\APACrefatitle {Particle methods for change detection, system
  identification, and control} {Particle methods for change detection, system
  identification, and control}.{\BBCQ}
\newblock
\APACjournalVolNumPages{Proceedings of the IEEE}{92}{3}{423--438}.
\PrintBackRefs{\CurrentBib}

\bibitem [\protect \citeauthoryear {%
Bagnara%
, Hill%
\BCBL {}\ \BBA {} Zaffanella%
}{%
Bagnara%
\ \protect \BOthers {.}}{%
{\protect \APACyear {2008}}%
}]{%
bagnara2008parma}
\APACinsertmetastar {%
bagnara2008parma}%
\begin{APACrefauthors}%
Bagnara, R.%
, Hill, P\BPBI M.%
\BCBL {}\ \BBA {} Zaffanella, E.%
\end{APACrefauthors}%
\unskip\
\newblock
\APACrefYearMonthDay{2008}{}{}.
\newblock
{\BBOQ}\APACrefatitle {The Parma Polyhedra Library: Toward a complete set of
  numerical abstractions for the analysis and verification of hardware and
  software systems} {The parma polyhedra library: Toward a complete set of
  numerical abstractions for the analysis and verification of hardware and
  software systems}.{\BBCQ}
\newblock
\APACjournalVolNumPages{Science of Computer Programming}{72}{1-2}{3--21}.
\PrintBackRefs{\CurrentBib}

\bibitem [\protect \citeauthoryear {%
Baumert%
\ \protect \BOthers {.}}{%
Baumert%
\ \protect \BOthers {.}}{%
{\protect \APACyear {2009}}%
}]{%
baumert2009discrete}
\APACinsertmetastar {%
baumert2009discrete}%
\begin{APACrefauthors}%
Baumert, S.%
, Ghate, A.%
, Kiatsupaibul, S.%
, Shen, Y.%
, Smith, R\BPBI L.%
\BCBL {}\ \BBA {} Zabinsky, Z\BPBI B.%
\end{APACrefauthors}%
\unskip\
\newblock
\APACrefYearMonthDay{2009}{}{}.
\newblock
{\BBOQ}\APACrefatitle {Discrete hit-and-run for sampling points from arbitrary
  distributions over subsets of integer hyperrectangles} {Discrete hit-and-run
  for sampling points from arbitrary distributions over subsets of integer
  hyperrectangles}.{\BBCQ}
\newblock
\APACjournalVolNumPages{Operations Research}{57}{3}{727--739}.
\PrintBackRefs{\CurrentBib}

\bibitem [\protect \citeauthoryear {%
Becchi%
\ \BBA {} Zaffanella%
}{%
Becchi%
\ \BBA {} Zaffanella%
}{%
{\protect \APACyear {2018}}%
}]{%
becchi2018efficient}
\APACinsertmetastar {%
becchi2018efficient}%
\begin{APACrefauthors}%
Becchi, A.%
\BCBT {}\ \BBA {} Zaffanella, E.%
\end{APACrefauthors}%
\unskip\
\newblock
\APACrefYearMonthDay{2018}{}{}.
\newblock
{\BBOQ}\APACrefatitle {An efficient abstract domain for not necessarily closed
  polyhedra} {An efficient abstract domain for not necessarily closed
  polyhedra}.{\BBCQ}
\newblock
\BIn{} \APACrefbtitle {International Static Analysis Symposium} {International
  static analysis symposium}\ (\BPGS\ 146--165).
\PrintBackRefs{\CurrentBib}

\bibitem [\protect \citeauthoryear {%
Carrassi%
, Bocquet%
, Bertino%
\BCBL {}\ \BBA {} Evensen%
}{%
Carrassi%
\ \protect \BOthers {.}}{%
{\protect \APACyear {2018}}%
}]{%
carrassi2018data}
\APACinsertmetastar {%
carrassi2018data}%
\begin{APACrefauthors}%
Carrassi, A.%
, Bocquet, M.%
, Bertino, L.%
\BCBL {}\ \BBA {} Evensen, G.%
\end{APACrefauthors}%
\unskip\
\newblock
\APACrefYearMonthDay{2018}{}{}.
\newblock
{\BBOQ}\APACrefatitle {Data assimilation in the geosciences: An overview of
  methods, issues, and perspectives} {Data assimilation in the geosciences: An
  overview of methods, issues, and perspectives}.{\BBCQ}
\newblock
\APACjournalVolNumPages{Wiley Interdisciplinary Reviews: Climate
  Change}{9}{5}{e535}.
\PrintBackRefs{\CurrentBib}

\bibitem [\protect \citeauthoryear {%
Chatterjee%
\ \BBA {} Diaconis%
}{%
Chatterjee%
\ \BBA {} Diaconis%
}{%
{\protect \APACyear {2018}}%
}]{%
chatterjee2018sample}
\APACinsertmetastar {%
chatterjee2018sample}%
\begin{APACrefauthors}%
Chatterjee, S.%
\BCBT {}\ \BBA {} Diaconis, P.%
\end{APACrefauthors}%
\unskip\
\newblock
\APACrefYearMonthDay{2018}{}{}.
\newblock
{\BBOQ}\APACrefatitle {The sample size required in importance sampling} {The
  sample size required in importance sampling}.{\BBCQ}
\newblock
\APACjournalVolNumPages{The Annals of Applied Probability}{28}{2}{1099--1135}.
\PrintBackRefs{\CurrentBib}

\bibitem [\protect \citeauthoryear {%
Clay%
, Kieu%
, Ward%
, Heppenstall%
\BCBL {}\ \BBA {} Malleson%
}{%
Clay%
\ \protect \BOthers {.}}{%
{\protect \APACyear {2020}}%
}]{%
clay_realtime_2020}
\APACinsertmetastar {%
clay_realtime_2020}%
\begin{APACrefauthors}%
Clay, R.%
, Kieu, L\BHBI M.%
, Ward, J\BPBI A.%
, Heppenstall, A.%
\BCBL {}\ \BBA {} Malleson, N.%
\end{APACrefauthors}%
\unskip\
\newblock
\APACrefYearMonthDay{2020}{}{}.
\newblock
{\BBOQ}\APACrefatitle {Towards {{Real}}-{{Time Crowd Simulation Under
  Uncertainty Using}} an {{Agent}}-{{Based Model}} and an {{Unscented Kalman
  Filter}}} {Towards {{Real}}-{{Time Crowd Simulation Under Uncertainty Using}}
  an {{Agent}}-{{Based Model}} and an {{Unscented Kalman Filter}}}.{\BBCQ}
\newblock
\BIn{} Y.~Demazeau, T.~Holvoet, J\BPBI M.~Corchado\BCBL {}\ \BBA {}
  S.~Costantini\ (\BEDS), \APACrefbtitle {Advances in {{Practical
  Applications}} of {{Agents}}, {{Multi}}-{{Agent Systems}}, and
  {{Trustworthiness}}. {{The PAAMS Collection}}} {Advances in {{Practical
  Applications}} of {{Agents}}, {{Multi}}-{{Agent Systems}}, and
  {{Trustworthiness}}. {{The PAAMS Collection}}}\ (\BVOL\ 12092, \BPGS\
  68--79).
\newblock
\APACaddressPublisher{{Cham}}{{Springer International Publishing}}.
\newblock
\begin{APACrefDOI} \doi{10.1007/978-3-030-49778-1_6} \end{APACrefDOI}
\PrintBackRefs{\CurrentBib}

\bibitem [\protect \citeauthoryear {%
Conforti%
, Cornu{\'e}jols%
\BCBL {}\ \BBA {} Zambelli%
}{%
Conforti%
\ \protect \BOthers {.}}{%
{\protect \APACyear {2010}}%
}]{%
conforti2010polyhedral}
\APACinsertmetastar {%
conforti2010polyhedral}%
\begin{APACrefauthors}%
Conforti, M.%
, Cornu{\'e}jols, G.%
\BCBL {}\ \BBA {} Zambelli, G.%
\end{APACrefauthors}%
\unskip\
\newblock
\APACrefYearMonthDay{2010}{}{}.
\newblock
{\BBOQ}\APACrefatitle {Polyhedral approaches to mixed integer linear
  programming} {Polyhedral approaches to mixed integer linear
  programming}.{\BBCQ}
\newblock
\BIn{} \APACrefbtitle {50 years of integer programming 1958-2008} {50 years of
  integer programming 1958-2008}\ (\BPGS\ 343--385).
\newblock
\APACaddressPublisher{}{Springer}.
\PrintBackRefs{\CurrentBib}

\bibitem [\protect \citeauthoryear {%
Cousot%
\ \BBA {} Cousot%
}{%
Cousot%
\ \BBA {} Cousot%
}{%
{\protect \APACyear {1977}}%
}]{%
cousot1977abstract}
\APACinsertmetastar {%
cousot1977abstract}%
\begin{APACrefauthors}%
Cousot, P.%
\BCBT {}\ \BBA {} Cousot, R.%
\end{APACrefauthors}%
\unskip\
\newblock
\APACrefYearMonthDay{1977}{}{}.
\newblock
{\BBOQ}\APACrefatitle {Abstract interpretation: a unified lattice model for
  static analysis of programs by construction or approximation of fixpoints}
  {Abstract interpretation: a unified lattice model for static analysis of
  programs by construction or approximation of fixpoints}.{\BBCQ}
\newblock
\BIn{} \APACrefbtitle {Proceedings of the 4th ACM SIGACT-SIGPLAN symposium on
  Principles of programming languages} {Proceedings of the 4th acm
  sigact-sigplan symposium on principles of programming languages}\ (\BPGS\
  238--252).
\PrintBackRefs{\CurrentBib}

\bibitem [\protect \citeauthoryear {%
Cousot%
\ \BBA {} Halbwachs%
}{%
Cousot%
\ \BBA {} Halbwachs%
}{%
{\protect \APACyear {1978}}%
}]{%
cousot1978automatic}
\APACinsertmetastar {%
cousot1978automatic}%
\begin{APACrefauthors}%
Cousot, P.%
\BCBT {}\ \BBA {} Halbwachs, N.%
\end{APACrefauthors}%
\unskip\
\newblock
\APACrefYearMonthDay{1978}{}{}.
\newblock
{\BBOQ}\APACrefatitle {Automatic discovery of linear restraints among variables
  of a program} {Automatic discovery of linear restraints among variables of a
  program}.{\BBCQ}
\newblock
\BIn{} \APACrefbtitle {Proceedings of the 5th ACM SIGACT-SIGPLAN symposium on
  Principles of programming languages} {Proceedings of the 5th acm
  sigact-sigplan symposium on principles of programming languages}\ (\BPGS\
  84--96).
\PrintBackRefs{\CurrentBib}

\bibitem [\protect \citeauthoryear {%
Douc%
\ \BBA {} Capp{\'e}%
}{%
Douc%
\ \BBA {} Capp{\'e}%
}{%
{\protect \APACyear {2005}}%
}]{%
douc2005comparison}
\APACinsertmetastar {%
douc2005comparison}%
\begin{APACrefauthors}%
Douc, R.%
\BCBT {}\ \BBA {} Capp{\'e}, O.%
\end{APACrefauthors}%
\unskip\
\newblock
\APACrefYearMonthDay{2005}{}{}.
\newblock
{\BBOQ}\APACrefatitle {Comparison of resampling schemes for particle filtering}
  {Comparison of resampling schemes for particle filtering}.{\BBCQ}
\newblock
\BIn{} \APACrefbtitle {ISPA 2005. Proceedings of the 4th International
  Symposium on Image and Signal Processing and Analysis, 2005.} {Ispa 2005.
  proceedings of the 4th international symposium on image and signal processing
  and analysis, 2005.}\ (\BPGS\ 64--69).
\PrintBackRefs{\CurrentBib}

\bibitem [\protect \citeauthoryear {%
Doucet%
, Johansen%
\BCBL {}\ \protect \BOthers {.}}{%
Doucet%
\ \protect \BOthers {.}}{%
{\protect \APACyear {2009}}%
}]{%
doucet2009tutorial}
\APACinsertmetastar {%
doucet2009tutorial}%
\begin{APACrefauthors}%
Doucet, A.%
, Johansen, A\BPBI M.%
\BCBL {}\ \BOthersPeriod {.}\end{APACrefauthors}%
\unskip\
\newblock
\APACrefYearMonthDay{2009}{}{}.
\newblock
{\BBOQ}\APACrefatitle {A tutorial on particle filtering and smoothing: Fifteen
  years later} {A tutorial on particle filtering and smoothing: Fifteen years
  later}.{\BBCQ}
\newblock
\APACjournalVolNumPages{Handbook of nonlinear filtering}{12}{656-704}{3}.
\PrintBackRefs{\CurrentBib}

\bibitem [\protect \citeauthoryear {%
Evensen%
}{%
Evensen%
}{%
{\protect \APACyear {2003}}%
}]{%
evensen2003ensemble}
\APACinsertmetastar {%
evensen2003ensemble}%
\begin{APACrefauthors}%
Evensen, G.%
\end{APACrefauthors}%
\unskip\
\newblock
\APACrefYearMonthDay{2003}{}{}.
\newblock
{\BBOQ}\APACrefatitle {The ensemble Kalman filter: Theoretical formulation and
  practical implementation} {The ensemble kalman filter: Theoretical
  formulation and practical implementation}.{\BBCQ}
\newblock
\APACjournalVolNumPages{Ocean dynamics}{53}{4}{343--367}.
\PrintBackRefs{\CurrentBib}

\bibitem [\protect \citeauthoryear {%
Finke%
, Doucet%
\BCBL {}\ \BBA {} Johansen%
}{%
Finke%
\ \protect \BOthers {.}}{%
{\protect \APACyear {2020}}%
}]{%
finke2020limit}
\APACinsertmetastar {%
finke2020limit}%
\begin{APACrefauthors}%
Finke, A.%
, Doucet, A.%
\BCBL {}\ \BBA {} Johansen, A\BPBI M.%
\end{APACrefauthors}%
\unskip\
\newblock
\APACrefYearMonthDay{2020}{}{}.
\newblock
{\BBOQ}\APACrefatitle {Limit theorems for sequential MCMC methods} {Limit
  theorems for sequential mcmc methods}.{\BBCQ}
\newblock
\APACjournalVolNumPages{Advances in Applied Probability}{52}{2}{377--403}.
\PrintBackRefs{\CurrentBib}

\bibitem [\protect \citeauthoryear {%
Fukuda%
}{%
Fukuda%
}{%
{\protect \APACyear {2020}}%
}]{%
fukuda2020polyhedral}
\APACinsertmetastar {%
fukuda2020polyhedral}%
\begin{APACrefauthors}%
Fukuda, K.%
\end{APACrefauthors}%
\unskip\
\newblock
\APACrefYearMonthDay{2020}{}{}.
\newblock
{\BBOQ}\APACrefatitle {Polyhedral computation} {Polyhedral computation}.{\BBCQ}
\newblock

\newblock
\begin{APACrefDOI} \doi{https://doi.org/10.3929/ethz-b-000426218}
  \end{APACrefDOI}
\PrintBackRefs{\CurrentBib}

\bibitem [\protect \citeauthoryear {%
Gelman%
, Carlin%
, Stern%
\BCBL {}\ \BBA {} Rubin%
}{%
Gelman%
\ \protect \BOthers {.}}{%
{\protect \APACyear {2013}}%
}]{%
gelman2013bayesian}
\APACinsertmetastar {%
gelman2013bayesian}%
\begin{APACrefauthors}%
Gelman, A.%
, Carlin, J\BPBI B.%
, Stern, H\BPBI S.%
\BCBL {}\ \BBA {} Rubin, D\BPBI B.%
\end{APACrefauthors}%
\unskip\
\newblock
\APACrefYear{2013}.
\newblock
\APACrefbtitle {Bayesian data analysis} {Bayesian data analysis}\
  (\PrintOrdinal{3}\ \BEd).
\newblock
\begin{APACrefURL} [{06/01/22}]\url{http://www.stat.columbia.edu/~gelman/book/}
  \end{APACrefURL}
\PrintBackRefs{\CurrentBib}

\bibitem [\protect \citeauthoryear {%
Gelman%
\ \BBA {} Rubin%
}{%
Gelman%
\ \BBA {} Rubin%
}{%
{\protect \APACyear {1992}}%
}]{%
gelman1992inference}
\APACinsertmetastar {%
gelman1992inference}%
\begin{APACrefauthors}%
Gelman, A.%
\BCBT {}\ \BBA {} Rubin, D\BPBI B.%
\end{APACrefauthors}%
\unskip\
\newblock
\APACrefYearMonthDay{1992}{}{}.
\newblock
{\BBOQ}\APACrefatitle {Inference from iterative simulation using multiple
  sequences} {Inference from iterative simulation using multiple
  sequences}.{\BBCQ}
\newblock
\APACjournalVolNumPages{Statistical science}{7}{4}{457--472}.
\PrintBackRefs{\CurrentBib}

\bibitem [\protect \citeauthoryear {%
Gilks%
\ \BBA {} Berzuini%
}{%
Gilks%
\ \BBA {} Berzuini%
}{%
{\protect \APACyear {2001}}%
}]{%
gilks2001following}
\APACinsertmetastar {%
gilks2001following}%
\begin{APACrefauthors}%
Gilks, W\BPBI R.%
\BCBT {}\ \BBA {} Berzuini, C.%
\end{APACrefauthors}%
\unskip\
\newblock
\APACrefYearMonthDay{2001}{}{}.
\newblock
{\BBOQ}\APACrefatitle {Following a moving target—Monte Carlo inference for
  dynamic Bayesian models} {Following a moving target—monte carlo inference
  for dynamic bayesian models}.{\BBCQ}
\newblock
\APACjournalVolNumPages{Journal of the Royal Statistical Society: Series B
  (Statistical Methodology)}{63}{1}{127--146}.
\PrintBackRefs{\CurrentBib}

\bibitem [\protect \citeauthoryear {%
Gordon%
, Salmond%
\BCBL {}\ \BBA {} Smith%
}{%
Gordon%
\ \protect \BOthers {.}}{%
{\protect \APACyear {1993}}%
}]{%
gordon1993novel}
\APACinsertmetastar {%
gordon1993novel}%
\begin{APACrefauthors}%
Gordon, N\BPBI J.%
, Salmond, D\BPBI J.%
\BCBL {}\ \BBA {} Smith, A\BPBI F.%
\end{APACrefauthors}%
\unskip\
\newblock
\APACrefYearMonthDay{1993}{}{}.
\newblock
{\BBOQ}\APACrefatitle {Novel approach to nonlinear/non-Gaussian Bayesian state
  estimation} {Novel approach to nonlinear/non-gaussian bayesian state
  estimation}.{\BBCQ}
\newblock
\BIn{} \APACrefbtitle {IEE Proceedings F-radar and signal processing} {Iee
  proceedings f-radar and signal processing}\ (\BVOL~140, \BPGS\ 107--113).
\PrintBackRefs{\CurrentBib}

\bibitem [\protect \citeauthoryear {%
Gurfinkel%
\ \BBA {} Navas%
}{%
Gurfinkel%
\ \BBA {} Navas%
}{%
{\protect \APACyear {2021}}%
}]{%
GN2021}
\APACinsertmetastar {%
GN2021}%
\begin{APACrefauthors}%
Gurfinkel, A.%
\BCBT {}\ \BBA {} Navas, J\BPBI A.%
\end{APACrefauthors}%
\unskip\
\newblock
\APACrefYearMonthDay{2021}{}{}.
\newblock
{\BBOQ}\APACrefatitle {Abstract Interpretation of {LLVM} with a Region-based
  Memory Model} {Abstract interpretation of {LLVM} with a region-based memory
  model}.{\BBCQ}
\newblock
\BIn{} \APACrefbtitle {Software Verification - 13th International Conference,
  {VSTTE} 2021, and 14th International Workshop, {NSV} 2021, October 18-19,
  2021, Revised Selected Papers.} {Software verification - 13th international
  conference, {VSTTE} 2021, and 14th international workshop, {NSV} 2021,
  october 18-19, 2021, revised selected papers.}
\PrintBackRefs{\CurrentBib}

\bibitem [\protect \citeauthoryear {%
Halbwachs%
, Proy%
\BCBL {}\ \BBA {} Roumanoff%
}{%
Halbwachs%
\ \protect \BOthers {.}}{%
{\protect \APACyear {1997}}%
}]{%
halbwachs1997verification}
\APACinsertmetastar {%
halbwachs1997verification}%
\begin{APACrefauthors}%
Halbwachs, N.%
, Proy, Y\BHBI E.%
\BCBL {}\ \BBA {} Roumanoff, P.%
\end{APACrefauthors}%
\unskip\
\newblock
\APACrefYearMonthDay{1997}{}{}.
\newblock
{\BBOQ}\APACrefatitle {Verification of real-time systems using linear relation
  analysis} {Verification of real-time systems using linear relation
  analysis}.{\BBCQ}
\newblock
\APACjournalVolNumPages{Formal Methods in System Design}{11}{2}{157--185}.
\PrintBackRefs{\CurrentBib}

\bibitem [\protect \citeauthoryear {%
Han%
\ \BBA {} Carlin%
}{%
Han%
\ \BBA {} Carlin%
}{%
{\protect \APACyear {2001}}%
}]{%
han2001markov}
\APACinsertmetastar {%
han2001markov}%
\begin{APACrefauthors}%
Han, C.%
\BCBT {}\ \BBA {} Carlin, B\BPBI P.%
\end{APACrefauthors}%
\unskip\
\newblock
\APACrefYearMonthDay{2001}{}{}.
\newblock
{\BBOQ}\APACrefatitle {Markov chain Monte Carlo methods for computing Bayes
  factors: A comparative review} {Markov chain monte carlo methods for
  computing bayes factors: A comparative review}.{\BBCQ}
\newblock
\APACjournalVolNumPages{Journal of the American Statistical
  Association}{96}{455}{1122--1132}.
\PrintBackRefs{\CurrentBib}

\bibitem [\protect \citeauthoryear {%
Henry%
, Monniaux%
\BCBL {}\ \BBA {} Moy%
}{%
Henry%
\ \protect \BOthers {.}}{%
{\protect \APACyear {2012}}%
}]{%
henry2012pagai}
\APACinsertmetastar {%
henry2012pagai}%
\begin{APACrefauthors}%
Henry, J.%
, Monniaux, D.%
\BCBL {}\ \BBA {} Moy, M.%
\end{APACrefauthors}%
\unskip\
\newblock
\APACrefYearMonthDay{2012}{}{}.
\newblock
{\BBOQ}\APACrefatitle {Pagai: A path sensitive static analyser} {Pagai: A path
  sensitive static analyser}.{\BBCQ}
\newblock
\APACjournalVolNumPages{Electronic Notes in Theoretical Computer
  Science}{289}{}{15--25}.
\PrintBackRefs{\CurrentBib}

\bibitem [\protect \citeauthoryear {%
Jeannet%
\ \BBA {} Min{\'e}%
}{%
Jeannet%
\ \BBA {} Min{\'e}%
}{%
{\protect \APACyear {2009}}%
}]{%
jeannet2009apron}
\APACinsertmetastar {%
jeannet2009apron}%
\begin{APACrefauthors}%
Jeannet, B.%
\BCBT {}\ \BBA {} Min{\'e}, A.%
\end{APACrefauthors}%
\unskip\
\newblock
\APACrefYearMonthDay{2009}{}{}.
\newblock
{\BBOQ}\APACrefatitle {Apron: A library of numerical abstract domains for
  static analysis} {Apron: A library of numerical abstract domains for static
  analysis}.{\BBCQ}
\newblock
\BIn{} \APACrefbtitle {International Conference on Computer Aided Verification}
  {International conference on computer aided verification}\ (\BPGS\ 661--667).
\PrintBackRefs{\CurrentBib}

\bibitem [\protect \citeauthoryear {%
Kalnay%
}{%
Kalnay%
}{%
{\protect \APACyear {2003}}%
}]{%
kalnay_atmospheric_2003}
\APACinsertmetastar {%
kalnay_atmospheric_2003}%
\begin{APACrefauthors}%
Kalnay, E.%
\end{APACrefauthors}%
\unskip\
\newblock
\APACrefYear{2003}.
\newblock
\APACrefbtitle {Atmospheric {{Modeling}}, {{Data Assimilation}} and
  {{Predictability}}} {Atmospheric {{Modeling}}, {{Data Assimilation}} and
  {{Predictability}}}.
\newblock
\APACaddressPublisher{}{{Cambridge University Press}}.
\PrintBackRefs{\CurrentBib}

\bibitem [\protect \citeauthoryear {%
Khan%
, Balch%
\BCBL {}\ \BBA {} Dellaert%
}{%
Khan%
\ \protect \BOthers {.}}{%
{\protect \APACyear {2003}}%
}]{%
khan2003efficient}
\APACinsertmetastar {%
khan2003efficient}%
\begin{APACrefauthors}%
Khan, Z.%
, Balch, T.%
\BCBL {}\ \BBA {} Dellaert, F.%
\end{APACrefauthors}%
\unskip\
\newblock
\APACrefYearMonthDay{2003}{}{}.
\newblock
{\BBOQ}\APACrefatitle {Efficient particle filter-based tracking of multiple
  interacting targets using an MRF-based motion model} {Efficient particle
  filter-based tracking of multiple interacting targets using an mrf-based
  motion model}.{\BBCQ}
\newblock
\BIn{} \APACrefbtitle {Proceedings 2003 IEEE/RSJ International Conference on
  Intelligent Robots and Systems (IROS 2003)(Cat. No. 03CH37453)} {Proceedings
  2003 ieee/rsj international conference on intelligent robots and systems
  (iros 2003)(cat. no. 03ch37453)}\ (\BVOL~1, \BPGS\ 254--259).
\PrintBackRefs{\CurrentBib}

\bibitem [\protect \citeauthoryear {%
Kieu%
, Malleson%
\BCBL {}\ \BBA {} Heppenstall%
}{%
Kieu%
\ \protect \BOthers {.}}{%
{\protect \APACyear {2020}}%
}]{%
kieu_dealing_2020}
\APACinsertmetastar {%
kieu_dealing_2020}%
\begin{APACrefauthors}%
Kieu, L\BHBI M.%
, Malleson, N.%
\BCBL {}\ \BBA {} Heppenstall, A.%
\end{APACrefauthors}%
\unskip\
\newblock
\APACrefYearMonthDay{2020}{}{}.
\newblock
{\BBOQ}\APACrefatitle {Dealing with Uncertainty in Agent-Based Models for
  Short-Term Predictions} {Dealing with uncertainty in agent-based models for
  short-term predictions}.{\BBCQ}
\newblock
\APACjournalVolNumPages{Royal Society Open Science}{7}{1}{191074}.
\newblock
\begin{APACrefDOI} \doi{10.1098/rsos.191074} \end{APACrefDOI}
\PrintBackRefs{\CurrentBib}

\bibitem [\protect \citeauthoryear {%
Kirkpatrick%
, Gelatt%
\BCBL {}\ \BBA {} Vecchi%
}{%
Kirkpatrick%
\ \protect \BOthers {.}}{%
{\protect \APACyear {1983}}%
}]{%
kirkpatrick1983optimization}
\APACinsertmetastar {%
kirkpatrick1983optimization}%
\begin{APACrefauthors}%
Kirkpatrick, S.%
, Gelatt, C\BPBI D.%
\BCBL {}\ \BBA {} Vecchi, M\BPBI P.%
\end{APACrefauthors}%
\unskip\
\newblock
\APACrefYearMonthDay{1983}{}{}.
\newblock
{\BBOQ}\APACrefatitle {Optimization by simulated annealing} {Optimization by
  simulated annealing}.{\BBCQ}
\newblock
\APACjournalVolNumPages{science}{220}{4598}{671--680}.
\PrintBackRefs{\CurrentBib}

\bibitem [\protect \citeauthoryear {%
Lewis%
, Lakshmivarahan%
\BCBL {}\ \BBA {} Dhall%
}{%
Lewis%
\ \protect \BOthers {.}}{%
{\protect \APACyear {2006}}%
}]{%
lewis_dynamic_2006}
\APACinsertmetastar {%
lewis_dynamic_2006}%
\begin{APACrefauthors}%
Lewis, J\BPBI M.%
, Lakshmivarahan, S.%
\BCBL {}\ \BBA {} Dhall, S.%
\end{APACrefauthors}%
\unskip\
\newblock
\APACrefYear{2006}.
\newblock
\APACrefbtitle {Dynamic {{Data Assimilation}}: {{A Least Squares Approach}}}
  {Dynamic {{Data Assimilation}}: {{A Least Squares Approach}}}.
\newblock
\APACaddressPublisher{{Cambridge}}{{Cambridge University Press}}.
\PrintBackRefs{\CurrentBib}

\bibitem [\protect \citeauthoryear {%
Li%
, Sun%
, Sattar%
\BCBL {}\ \BBA {} Corchado%
}{%
Li%
\ \protect \BOthers {.}}{%
{\protect \APACyear {2014}}%
}]{%
li2014fight}
\APACinsertmetastar {%
li2014fight}%
\begin{APACrefauthors}%
Li, T.%
, Sun, S.%
, Sattar, T\BPBI P.%
\BCBL {}\ \BBA {} Corchado, J\BPBI M.%
\end{APACrefauthors}%
\unskip\
\newblock
\APACrefYearMonthDay{2014}{}{}.
\newblock
{\BBOQ}\APACrefatitle {Fight sample degeneracy and impoverishment in particle
  filters: A review of intelligent approaches} {Fight sample degeneracy and
  impoverishment in particle filters: A review of intelligent
  approaches}.{\BBCQ}
\newblock
\APACjournalVolNumPages{Expert Systems with applications}{41}{8}{3944--3954}.
\PrintBackRefs{\CurrentBib}

\bibitem [\protect \citeauthoryear {%
Liao%
\ \BBA {} Barooah%
}{%
Liao%
\ \BBA {} Barooah%
}{%
{\protect \APACyear {2010}}%
}]{%
liao2010integrated}
\APACinsertmetastar {%
liao2010integrated}%
\begin{APACrefauthors}%
Liao, C.%
\BCBT {}\ \BBA {} Barooah, P.%
\end{APACrefauthors}%
\unskip\
\newblock
\APACrefYearMonthDay{2010}{}{}.
\newblock
{\BBOQ}\APACrefatitle {An integrated approach to occupancy modeling and
  estimation in commercial buildings} {An integrated approach to occupancy
  modeling and estimation in commercial buildings}.{\BBCQ}
\newblock
\BIn{} \APACrefbtitle {Proceedings of the 2010 American control conference}
  {Proceedings of the 2010 american control conference}\ (\BPGS\ 3130--3135).
\PrintBackRefs{\CurrentBib}

\bibitem [\protect \citeauthoryear {%
Liu%
\ \BBA {} West%
}{%
Liu%
\ \BBA {} West%
}{%
{\protect \APACyear {2001}}%
}]{%
liu2001combined}
\APACinsertmetastar {%
liu2001combined}%
\begin{APACrefauthors}%
Liu, J.%
\BCBT {}\ \BBA {} West, M.%
\end{APACrefauthors}%
\unskip\
\newblock
\APACrefYearMonthDay{2001}{}{}.
\newblock
{\BBOQ}\APACrefatitle {Combined parameter and state estimation in
  simulation-based filtering} {Combined parameter and state estimation in
  simulation-based filtering}.{\BBCQ}
\newblock
\BIn{} \APACrefbtitle {Sequential Monte Carlo methods in practice} {Sequential
  monte carlo methods in practice}\ (\BPGS\ 197--223).
\newblock
\APACaddressPublisher{}{Springer}.
\PrintBackRefs{\CurrentBib}

\bibitem [\protect \citeauthoryear {%
Lloyd%
, Santitissadeekorn%
\BCBL {}\ \BBA {} Short%
}{%
Lloyd%
\ \protect \BOthers {.}}{%
{\protect \APACyear {2016}}%
}]{%
lloyd_exploring_2016}
\APACinsertmetastar {%
lloyd_exploring_2016}%
\begin{APACrefauthors}%
Lloyd, D\BPBI J\BPBI B.%
, Santitissadeekorn, N.%
\BCBL {}\ \BBA {} Short, M\BPBI B.%
\end{APACrefauthors}%
\unskip\
\newblock
\APACrefYearMonthDay{2016}{}{}.
\newblock
{\BBOQ}\APACrefatitle {Exploring Data Assimilation and Forecasting Issues for
  an Urban Crime Model} {Exploring data assimilation and forecasting issues for
  an urban crime model}.{\BBCQ}
\newblock
\APACjournalVolNumPages{European Journal of Applied Mathematics}{27}{Special
  Issue 03}{451--478}.
\newblock
\begin{APACrefDOI} \doi{10.1017/S0956792515000625} \end{APACrefDOI}
\PrintBackRefs{\CurrentBib}

\bibitem [\protect \citeauthoryear {%
Lueck%
, Rife%
, Swarup%
\BCBL {}\ \BBA {} Uddin%
}{%
Lueck%
\ \protect \BOthers {.}}{%
{\protect \APACyear {2019}}%
}]{%
lueck_who_2019}
\APACinsertmetastar {%
lueck_who_2019}%
\begin{APACrefauthors}%
Lueck, J.%
, Rife, J\BPBI H.%
, Swarup, S.%
\BCBL {}\ \BBA {} Uddin, N.%
\end{APACrefauthors}%
\unskip\
\newblock
\APACrefYearMonthDay{2019}{}{}.
\newblock
{\BBOQ}\APACrefatitle {Who {{Goes There}}? {{Using}} an {{Agent}}-Based
  {{Simulation}} for {{Tracking Population Movement}}} {Who {{Goes There}}?
  {{Using}} an {{Agent}}-based {{Simulation}} for {{Tracking Population
  Movement}}}.{\BBCQ}
\newblock
\BIn{} \APACrefbtitle {Winter {{Simulation Conference}}, {{Dec}} 8 - 11, 2019.}
  {Winter {{Simulation Conference}}, {{Dec}} 8 - 11, 2019.}
\newblock
\APACaddressPublisher{{National Harbor, MD, USA}}{}.
\PrintBackRefs{\CurrentBib}

\bibitem [\protect \citeauthoryear {%
Malleson%
\ \protect \BOthers {.}}{%
Malleson%
\ \protect \BOthers {.}}{%
{\protect \APACyear {2020}}%
}]{%
malleson_simulating_2020}
\APACinsertmetastar {%
malleson_simulating_2020}%
\begin{APACrefauthors}%
Malleson, N.%
, Minors, K.%
, Kieu, L\BHBI M.%
, Ward, J\BPBI A.%
, West, A.%
\BCBL {}\ \BBA {} Heppenstall, A.%
\end{APACrefauthors}%
\unskip\
\newblock
\APACrefYearMonthDay{2020}{}{}.
\newblock
{\BBOQ}\APACrefatitle {Simulating Crowds in Real Time with Agent-Based
  Modelling and a Particle Filter} {Simulating crowds in real time with
  agent-based modelling and a particle filter}.{\BBCQ}
\newblock
\APACjournalVolNumPages{Journal of Artificial Societies and Social
  Simulation}{23}{3}{3}.
\newblock
\begin{APACrefDOI} \doi{10.18564/jasss.4266} \end{APACrefDOI}
\PrintBackRefs{\CurrentBib}

\bibitem [\protect \citeauthoryear {%
Markowitz%
}{%
Markowitz%
}{%
{\protect \APACyear {1957}}%
}]{%
markowitz1957elimination}
\APACinsertmetastar {%
markowitz1957elimination}%
\begin{APACrefauthors}%
Markowitz, H\BPBI M.%
\end{APACrefauthors}%
\unskip\
\newblock
\APACrefYearMonthDay{1957}{}{}.
\newblock
{\BBOQ}\APACrefatitle {The elimination form of the inverse and its application
  to linear programming} {The elimination form of the inverse and its
  application to linear programming}.{\BBCQ}
\newblock
\APACjournalVolNumPages{Management Science}{3}{3}{255--269}.
\PrintBackRefs{\CurrentBib}

\bibitem [\protect \citeauthoryear {%
Maros%
}{%
Maros%
}{%
{\protect \APACyear {2002}}%
}]{%
maros2002computational}
\APACinsertmetastar {%
maros2002computational}%
\begin{APACrefauthors}%
Maros, I.%
\end{APACrefauthors}%
\unskip\
\newblock
\APACrefYear{2002}.
\newblock
\APACrefbtitle {Computational techniques of the simplex method} {Computational
  techniques of the simplex method}\ (\BVOL~61).
\newblock
\APACaddressPublisher{}{Springer Science \& Business Media}.
\PrintBackRefs{\CurrentBib}

\bibitem [\protect \citeauthoryear {%
Meel%
\ \protect \BOthers {.}}{%
Meel%
\ \protect \BOthers {.}}{%
{\protect \APACyear {2016}}%
}]{%
meel2016constrained}
\APACinsertmetastar {%
meel2016constrained}%
\begin{APACrefauthors}%
Meel, K\BPBI S.%
, Vardi, M\BPBI Y.%
, Chakraborty, S.%
, Fremont, D\BPBI J.%
, Seshia, S\BPBI A.%
, Fried, D.%
\BDBL {}Malik, S.%
\end{APACrefauthors}%
\unskip\
\newblock
\APACrefYearMonthDay{2016}{}{}.
\newblock
{\BBOQ}\APACrefatitle {Constrained sampling and counting: Universal hashing
  meets SAT solving} {Constrained sampling and counting: Universal hashing
  meets sat solving}.{\BBCQ}
\newblock
\BIn{} \APACrefbtitle {Workshops at the thirtieth AAAI conference on artificial
  intelligence.} {Workshops at the thirtieth aaai conference on artificial
  intelligence.}
\PrintBackRefs{\CurrentBib}

\bibitem [\protect \citeauthoryear {%
Mihelich%
, Dubrulle%
, Paillard%
, Kral%
\BCBL {}\ \BBA {} Faranda%
}{%
Mihelich%
\ \protect \BOthers {.}}{%
{\protect \APACyear {2018}}%
}]{%
mihelich2018maximum}
\APACinsertmetastar {%
mihelich2018maximum}%
\begin{APACrefauthors}%
Mihelich, M.%
, Dubrulle, B.%
, Paillard, D.%
, Kral, Q.%
\BCBL {}\ \BBA {} Faranda, D.%
\end{APACrefauthors}%
\unskip\
\newblock
\APACrefYearMonthDay{2018}{}{}.
\newblock
{\BBOQ}\APACrefatitle {Maximum kolmogorov-sinai entropy versus minimum mixing
  time in markov chains} {Maximum kolmogorov-sinai entropy versus minimum
  mixing time in markov chains}.{\BBCQ}
\newblock
\APACjournalVolNumPages{Journal of Statistical Physics}{170}{1}{62--68}.
\PrintBackRefs{\CurrentBib}

\bibitem [\protect \citeauthoryear {%
Motzkin%
, Raiffa%
, Thompson%
\BCBL {}\ \BBA {} Thrall%
}{%
Motzkin%
\ \protect \BOthers {.}}{%
{\protect \APACyear {1953}}%
}]{%
motzkin1953double}
\APACinsertmetastar {%
motzkin1953double}%
\begin{APACrefauthors}%
Motzkin, T\BPBI S.%
, Raiffa, H.%
, Thompson, G\BPBI L.%
\BCBL {}\ \BBA {} Thrall, R\BPBI M.%
\end{APACrefauthors}%
\unskip\
\newblock
\APACrefYearMonthDay{1953}{}{}.
\newblock
{\BBOQ}\APACrefatitle {The double description method} {The double description
  method}.{\BBCQ}
\newblock
\APACjournalVolNumPages{Contributions to the Theory of Games}{2}{28}{51--73}.
\PrintBackRefs{\CurrentBib}

\bibitem [\protect \citeauthoryear {%
Mu{\~n}oz-Pichardo%
, Pino-Mej{\'\i}as%
, Garc{\'\i}a-Heras%
, Ruiz-Mu{\~n}oz%
\BCBL {}\ \BBA {} Luz Gonz{\'a}lez-Regalado%
}{%
Mu{\~n}oz-Pichardo%
\ \protect \BOthers {.}}{%
{\protect \APACyear {2021}}%
}]{%
munoz2021multivariate}
\APACinsertmetastar {%
munoz2021multivariate}%
\begin{APACrefauthors}%
Mu{\~n}oz-Pichardo, J.%
, Pino-Mej{\'\i}as, R.%
, Garc{\'\i}a-Heras, J.%
, Ruiz-Mu{\~n}oz, F.%
\BCBL {}\ \BBA {} Luz Gonz{\'a}lez-Regalado, M.%
\end{APACrefauthors}%
\unskip\
\newblock
\APACrefYearMonthDay{2021}{}{}.
\newblock
{\BBOQ}\APACrefatitle {A multivariate Poisson regression model for count data}
  {A multivariate poisson regression model for count data}.{\BBCQ}
\newblock
\APACjournalVolNumPages{Journal of Applied Statistics}{48}{13-15}{2525--2541}.
\PrintBackRefs{\CurrentBib}

\bibitem [\protect \citeauthoryear {%
Newton%
\ \BBA {} Raftery%
}{%
Newton%
\ \BBA {} Raftery%
}{%
{\protect \APACyear {1994}}%
}]{%
newton1994approximate}
\APACinsertmetastar {%
newton1994approximate}%
\begin{APACrefauthors}%
Newton, M\BPBI A.%
\BCBT {}\ \BBA {} Raftery, A\BPBI E.%
\end{APACrefauthors}%
\unskip\
\newblock
\APACrefYearMonthDay{1994}{}{}.
\newblock
{\BBOQ}\APACrefatitle {Approximate Bayesian inference with the weighted
  likelihood bootstrap} {Approximate bayesian inference with the weighted
  likelihood bootstrap}.{\BBCQ}
\newblock
\APACjournalVolNumPages{Journal of the Royal Statistical Society: Series B
  (Methodological)}{56}{1}{3--26}.
\PrintBackRefs{\CurrentBib}

\bibitem [\protect \citeauthoryear {%
Schelling%
}{%
Schelling%
}{%
{\protect \APACyear {1971}}%
}]{%
schelling1971dynamic}
\APACinsertmetastar {%
schelling1971dynamic}%
\begin{APACrefauthors}%
Schelling, T\BPBI C.%
\end{APACrefauthors}%
\unskip\
\newblock
\APACrefYearMonthDay{1971}{}{}.
\newblock
{\BBOQ}\APACrefatitle {Dynamic models of segregation} {Dynamic models of
  segregation}.{\BBCQ}
\newblock
\APACjournalVolNumPages{Journal of mathematical sociology}{1}{2}{143--186}.
\PrintBackRefs{\CurrentBib}

\bibitem [\protect \citeauthoryear {%
Septier%
, Pang%
, Carmi%
\BCBL {}\ \BBA {} Godsill%
}{%
Septier%
\ \protect \BOthers {.}}{%
{\protect \APACyear {2009}}%
}]{%
septier2009mcmc}
\APACinsertmetastar {%
septier2009mcmc}%
\begin{APACrefauthors}%
Septier, F.%
, Pang, S\BPBI K.%
, Carmi, A.%
\BCBL {}\ \BBA {} Godsill, S.%
\end{APACrefauthors}%
\unskip\
\newblock
\APACrefYearMonthDay{2009}{}{}.
\newblock
{\BBOQ}\APACrefatitle {On MCMC-based particle methods for Bayesian filtering:
  Application to multitarget tracking} {On mcmc-based particle methods for
  bayesian filtering: Application to multitarget tracking}.{\BBCQ}
\newblock
\BIn{} \APACrefbtitle {3rd IEEE International Workshop on Computational
  Advances in Multi-Sensor Adaptive Processing (CAMSAP)} {3rd ieee
  international workshop on computational advances in multi-sensor adaptive
  processing (camsap)}\ (\BPGS\ 360--363).
\PrintBackRefs{\CurrentBib}

\bibitem [\protect \citeauthoryear {%
Sj{\"o}din%
\ \protect \BOthers {.}}{%
Sj{\"o}din%
\ \protect \BOthers {.}}{%
{\protect \APACyear {2009}}%
}]{%
nsjodin2009design}
\APACinsertmetastar {%
nsjodin2009design}%
\begin{APACrefauthors}%
Sj{\"o}din, J.%
, Pop, S.%
, Jagasia, H.%
, Grosser, T.%
, Pop, A.%
\BCBL {}\ \BOthersPeriod {.}\end{APACrefauthors}%
\unskip\
\newblock
\APACrefYearMonthDay{2009}{}{}.
\newblock
{\BBOQ}\APACrefatitle {Design of graphite and the polyhedral compilation
  package} {Design of graphite and the polyhedral compilation package}.{\BBCQ}
\newblock
\BIn{} \APACrefbtitle {GCC Developers’ Summit} {Gcc developers’ summit}\
  (\BPG~113).
\PrintBackRefs{\CurrentBib}

\bibitem [\protect \citeauthoryear {%
{\v{S}}tefankovi{\v{c}}%
, Vempala%
\BCBL {}\ \BBA {} Vigoda%
}{%
{\v{S}}tefankovi{\v{c}}%
\ \protect \BOthers {.}}{%
{\protect \APACyear {2009}}%
}]{%
stefankovic2009adaptive}
\APACinsertmetastar {%
stefankovic2009adaptive}%
\begin{APACrefauthors}%
{\v{S}}tefankovi{\v{c}}, D.%
, Vempala, S.%
\BCBL {}\ \BBA {} Vigoda, E.%
\end{APACrefauthors}%
\unskip\
\newblock
\APACrefYearMonthDay{2009}{}{}.
\newblock
{\BBOQ}\APACrefatitle {Adaptive simulated annealing: A near-optimal connection
  between sampling and counting} {Adaptive simulated annealing: A near-optimal
  connection between sampling and counting}.{\BBCQ}
\newblock
\APACjournalVolNumPages{Journal of the ACM (JACM)}{56}{3}{1--36}.
\PrintBackRefs{\CurrentBib}

\bibitem [\protect \citeauthoryear {%
Suhl%
\ \BBA {} Suhl%
}{%
Suhl%
\ \BBA {} Suhl%
}{%
{\protect \APACyear {1990}}%
}]{%
suhl1990computing}
\APACinsertmetastar {%
suhl1990computing}%
\begin{APACrefauthors}%
Suhl, U\BPBI H.%
\BCBT {}\ \BBA {} Suhl, L\BPBI M.%
\end{APACrefauthors}%
\unskip\
\newblock
\APACrefYearMonthDay{1990}{}{}.
\newblock
{\BBOQ}\APACrefatitle {Computing sparse LU factorizations for large-scale
  linear programming bases} {Computing sparse lu factorizations for large-scale
  linear programming bases}.{\BBCQ}
\newblock
\APACjournalVolNumPages{ORSA Journal on Computing}{2}{4}{325--335}.
\PrintBackRefs{\CurrentBib}

\bibitem [\protect \citeauthoryear {%
Talagrand%
}{%
Talagrand%
}{%
{\protect \APACyear {1997}}%
}]{%
talagrand_assimilation_1997}
\APACinsertmetastar {%
talagrand_assimilation_1997}%
\begin{APACrefauthors}%
Talagrand, O.%
\end{APACrefauthors}%
\unskip\
\newblock
\APACrefYearMonthDay{1997}{}{}.
\newblock
{\BBOQ}\APACrefatitle {Assimilation of {{Observations}}, an {{Introduction}}}
  {Assimilation of {{Observations}}, an {{Introduction}}}.{\BBCQ}
\newblock
\APACjournalVolNumPages{Journal of the Meteorological Society of Japan. Ser.
  II}{75}{1B}{191--209}.
\PrintBackRefs{\CurrentBib}

\bibitem [\protect \citeauthoryear {%
Tang%
}{%
Tang%
}{%
{\protect \APACyear {2019}}%
}]{%
tang2019data}
\APACinsertmetastar {%
tang2019data}%
\begin{APACrefauthors}%
Tang, D.%
\end{APACrefauthors}%
\unskip\
\newblock
\APACrefYearMonthDay{2019}{}{}.
\newblock
{\BBOQ}\APACrefatitle {Data assimilation in Agent-based models using creation
  and annihilation operators} {Data assimilation in agent-based models using
  creation and annihilation operators}.{\BBCQ}
\newblock
\APACjournalVolNumPages{arXiv preprint arXiv:1910.09442}{}{}{}.
\newblock
\begin{APACrefURL} \url{https://arxiv.org/abs/1910.09442} \end{APACrefURL}
\PrintBackRefs{\CurrentBib}

\bibitem [\protect \citeauthoryear {%
Tang%
}{%
Tang%
}{%
{\protect \APACyear {2022}}%
}]{%
TangMutableCategorical}
\APACinsertmetastar {%
TangMutableCategorical}%
\begin{APACrefauthors}%
Tang, D.%
\end{APACrefauthors}%
\unskip\
\newblock
\APACrefYearMonthDay{2022}{}{}.
\newblock
\APACrefbtitle {Mutable Categorical Distribution.} {Mutable categorical
  distribution.}
\newblock
\APACaddressPublisher{}{GitHub}.
\newblock
\begin{APACrefURL}
  [{06/01/22}]\url{https://github.com/danftang/MutableCategoricalDistribution}
  \end{APACrefURL}
\PrintBackRefs{\CurrentBib}

\bibitem [\protect \citeauthoryear {%
Wan%
, Van Der~Merwe%
\BCBL {}\ \BBA {} Haykin%
}{%
Wan%
\ \protect \BOthers {.}}{%
{\protect \APACyear {2001}}%
}]{%
wan2001unscented}
\APACinsertmetastar {%
wan2001unscented}%
\begin{APACrefauthors}%
Wan, E\BPBI A.%
, Van Der~Merwe, R.%
\BCBL {}\ \BBA {} Haykin, S.%
\end{APACrefauthors}%
\unskip\
\newblock
\APACrefYearMonthDay{2001}{}{}.
\newblock
{\BBOQ}\APACrefatitle {The unscented Kalman filter} {The unscented kalman
  filter}.{\BBCQ}
\newblock
\APACjournalVolNumPages{Kalman filtering and neural
  networks}{5}{2007}{221--280}.
\PrintBackRefs{\CurrentBib}

\bibitem [\protect \citeauthoryear {%
Wang%
\ \BBA {} Hu%
}{%
Wang%
\ \BBA {} Hu%
}{%
{\protect \APACyear {2015}}%
}]{%
wang_data_2015}
\APACinsertmetastar {%
wang_data_2015}%
\begin{APACrefauthors}%
Wang, M.%
\BCBT {}\ \BBA {} Hu, X.%
\end{APACrefauthors}%
\unskip\
\newblock
\APACrefYearMonthDay{2015}{}{}.
\newblock
{\BBOQ}\APACrefatitle {Data Assimilation in Agent Based Simulation of Smart
  Environments Using Particle Filters} {Data assimilation in agent based
  simulation of smart environments using particle filters}.{\BBCQ}
\newblock
\APACjournalVolNumPages{Simulation Modelling Practice and
  Theory}{56}{}{36--54}.
\newblock
\begin{APACrefDOI} \doi{10.1016/j.simpat.2015.05.001} \end{APACrefDOI}
\PrintBackRefs{\CurrentBib}

\bibitem [\protect \citeauthoryear {%
Ward%
, Evans%
\BCBL {}\ \BBA {} Malleson%
}{%
Ward%
\ \protect \BOthers {.}}{%
{\protect \APACyear {2016}}%
}]{%
ward_dynamic_2016}
\APACinsertmetastar {%
ward_dynamic_2016}%
\begin{APACrefauthors}%
Ward, J\BPBI A.%
, Evans, A\BPBI J.%
\BCBL {}\ \BBA {} Malleson, N\BPBI S.%
\end{APACrefauthors}%
\unskip\
\newblock
\APACrefYearMonthDay{2016}{}{}.
\newblock
{\BBOQ}\APACrefatitle {Dynamic Calibration of Agent-Based Models Using Data
  Assimilation} {Dynamic calibration of agent-based models using data
  assimilation}.{\BBCQ}
\newblock
\APACjournalVolNumPages{Royal Society Open Science}{3}{4}{}.
\newblock
\begin{APACrefDOI} \doi{10.1098/rsos.150703} \end{APACrefDOI}
\PrintBackRefs{\CurrentBib}

\end{thebibliography}

\end{document}